\documentstyle[11pt,epsfig]{article}

\newcommand{\reals}{\mbox{R}}
\newcommand{\complexes}{\mbox{C}}
\newcommand{\integers}{\mbox{Z}}

\newcommand{\beq}{\begin{equation}}
\newcommand{\eeq}{\end{equation}}
\newcommand{\beqa}{\begin{eqnarray}}
\newcommand{\eeqa}{\end{eqnarray}}
\newcommand{\bi}{\begin{itemize}}
\newcommand{\ei}{\end{itemize}}

\newcommand{\tr}{\mbox{tr}}

\newcommand{\fot}{\frac{1}{2}}

\def\a{\alpha}
\def\b{\beta}
\def\c{\gamma}
\def\d{\delta}
\def\e{\epsilon}
\def\z{\zeta}
\def\L{\Lambda}

\begin{document}

\begin{flushright}
MPI-Ph/93-94
\\
gr-qc/9312001
\\
December 1993
\end{flushright}
\vfill

\begin{center}{\Large\bf
Loop Representations
%\footnote{
%To appear in `Canonical Relativity from Classical to Quantum',
%ed.\ H. Friedrich (Springer Lecture Notes)}
%
}
\end{center}
\vfill

\begin{center}
Bernd Br\"ugmann
\end{center}
\bigskip\medskip

\begin{center}
{\small
\em Max-Planck-Institute of Physics, 80805 M\"unchen, Germany
\\
bruegman@iws170.mppmu.mpg.de
}
\end{center}
\vfill

\subsection*{Abstract}
The loop representation plays an important role in canonical quantum
gravity because loop variables allow a natural treatment of the
constraints. In these lectures we give an elementary introduction to (i)
the relevant history of loops in knot theory and gauge theory, (ii) the
loop representation of Maxwell theory, and (iii) the loop representation of
canonical quantum gravity.

\vspace*{\fill}
\newpage

\tableofcontents
\newpage

%%%%%%%%%%%%%%%%%%%%%%%%%%%%%%%%%%%%%%%%%%%%%%%%%%%%%%%%%%%%%%%%%%%%%%
\section{Introduction}

\subsection{Generalities}

The task of theoretical physics is to find an adequate mathematical
description of physical ideas, and there is always a tension between real
world experiments and the mathematical structures modelling them. In
quantum gravity this tension appears with a twist, since there is no direct
experimental evidence to be explained. Guided by the hypothesis that there
exists a unified description of nature, we try to find a theory that could
describe phenomena that belong at the same time to the domain of quantum
theory and general relativity.

It is generally agreed upon that such phenomena exist even though they have
not been observed. An example could be matter near the big bang
singularity, and predictions like that of Hawking radiation for quantum
fields near a black hole horizon are expected to eventually be valid
limiting cases of a full theory of quantum gravity. But there are widely
varying opinions on how one should go about constructing a theory of
quantum gravity.

If we examine the theories that we want to combine closely, they are not
flawless to begin with. One can for example argue that the measurement
process of quantum mechanics is not sufficiently explained, while classical
general relativity does not allow for quantum matter. In fact, we hope to
remedy such problems in the more complete theory of quantum gravity.

It is not the case that the lack of experimental data does not allow us to
pick the right theory out of a host of possibilities.  The most important
fact to remember about quantum gravity is that to date there does not exist
a single model for a quantum theory of general relativity that is (i)
self-consistent and (ii) contains as a special case a reasonable
approximation to the observed physical world. There exist many research
programs to construct quantum gravity, but all are incomplete even by their
own criteria. However, some interesting partial results have been
obtained in certain approaches, and here we will focus our attention on one
such approach.

We consider the program of canonical quantization of general relativity.
Although as incomplete as other approaches, there has been some recent
progress initiated by the discovery of a new set of canonical variables for
general relativity by Ashtekar \cite{As86}. The reason for our discussion
of loop representations is that the loop representation of Rovelli and
Smolin \cite{RoSm88} figures prominently in the quantum theory based on the
Ashtekar variables.

There are many facets to canonical quantum gravity. In particular, there is
much more to the program of canonical quantization starting from the
Ashtekar variables then the loop representation, and the Ashtekar variables
are in addition very interesting for the classical theory (see this
volume). The goal of this paper is to demonstrate that similarly there is
more to the concept of a loop representation than that it is a useful
technique in canonical quantum gravity. To this end we will try to paint a
coherent picture --- drawing on knot theory, gauge theory and canonical
quantum gravity --- of why loops find a natural place in canonical quantum
gravity.

There exist several excellent reviews of canonical quantum gravity, the
Ash\-te\-kar variables, and the loop representation (which we will point
out as we go along).  Our emphasis will be on a complete and up to date
development of the main ideas behind loop representations rather then on
technical details, in the hope that this way the {\em motivation} for the
loop representation becomes apparent and the reader may answer the
question, ``Why loops?''

\subsection{What we mean by canonical quantum gravity}

To put our discussion of quantum gravity into perspective, let us first
attach a few labels to what we mean and imply by the term 'canonical
quantum gravity' in these lectures. A much more thorough background to what
quantum gravity could mean can be found in the articles of Isham in this
volume.
\begin{enumerate}

\item Starting point is standard general relativity, i.e. the theory of a
Lorentzian metric $g_{\mu\nu}(x)$ on a four dimensional differentiable
manifold $\cal M$ defined by the Einstein-Hilbert action,
\beq
	S[g] = \int_{\cal M} d^4x \,\sqrt{- g} \, R,
\eeq
where $R$ is the Ricci scalar and $g$ the determinant of
$g_{\mu\nu}(x)$. There are all sorts of reasons why one might want to
consider discrete spaces instead of $\cal M$, or strings instead of points,
or actions modified by higher order curvature terms, or other
alterations. Whenever we refer to general relativity, we mean standard
general relativity.

\item
Even though there currently is a great deal interest in theories in
dimension unequal four, here we consider general relativity in four
spacetime dimensions if we do not state otherwise. In fact, one of the
reasons why loop representations are of interest is that they are tailored
to the physically observed number of four dimensions.

\item
Other approaches to quantum gravity, in particular path integral
quantization and quantum cosmology, are based on an Euclideanization. Since
Euclidean quantum gravity cannot in general be extended to Lorentzian
quantum gravity, it may be an advantage that canonical quantization and the
loop representation are Lorentzian (although they are simple to Euclideanize).

\item Since we consider a canonical formulation, we only allow spacetimes
that can be split into space and time, $\cal M = \Sigma \times \reals$. The
three-manifold $\Sigma$ is assumed to be compact for simplicity.
Such a choice for $\cal M$ excludes the possibility of topology change,
which can be accommodated, for example, in path integral quantization.

\item
Our goal is an inherently non-perturbative formulation, for which canonical
quantization is well-suited. This has to be contrasted with standard field
theoretic methods, which are based on perturbation theory using Feynman
diagrams obtained from path integrals. Simply put, perturbation theory for
general relativity fails at two loop, and the characteristica of general
relativity, e.g. diffeomorphism invariance, seem to make a non-perturbative
approach necessary.  Let us also point out that a non-perturbative,
canonical formulation looks and is in fact very different from perturbation
theory. Many of the unconventional features that we encounter in the loop
representation are not specific to the loop representation but to the
canonical approach in general.

\item
We consider gravity without matter, only at one point will we introduce a
cosmological constant. Matter can be incorporated into the loop
representation, but we will only briefly comment on it in section 5.1. The
attitude in particle physics is that gravity can be treated by the same
methods that are succesful for the other interactions, the only difference
being the energy scale. The most elegant scheme along these lines is
certainly string theory, although gravity is introduced via gravitons,
which is a concept from perturbation theory. The presence of matter is
thought to be essential. A common philosophy among relativists is that we
can learn something about the deep conceptual issues in quantum gravity
already in the absence of matter. Matter is thought to cloud some of these
issues. But for questions like the issue of time in quantum gravity, we
expect that we do have to include matter.

\item
Canonical quantization of the type considered here
is a very conservative approach. No new structures are postulated, like
strings or supersymmetry. The idea is to use new techniques rather than new
concepts as long as they are not forced upon us, i.e.\ to stay as close as
possible to conventional quantum mechanics.

\item A loop is a map from the circle into a manifold. While mathematically
the same objects as closed strings in string theory or loops in Feynman
diagrams, loops play a completely different role in the loop
representation. Also, their algebraic structure is not what is called a
loop group, although a group of loops can be defined.

\end{enumerate}

\subsection{Outline}

Loop representations have their origin in gauge theories, in fact, any
Yang-Mills theory admits formally a loop representation. Let us summarize
how a connection to general relativity is established.  Canonical
Yang-Mills theory is characterized by a gauge constraint, $G$, while
general relativity in the usual metric variables is characterized by a
diffeomorphism constraint $D$, which generates three-dimensional
diffeomorphisms in $\Sigma$ and the Hamiltonian constraint $H$, which
essentially specifies how $\Sigma$ is imbedded in $\cal M$.  In the
canonically quantized theory the constraints are imposed as operator
equations on the states $\psi$.

While quantum gravity in terms of the metric variables and Yang-Mills
theory in terms of a connection are disjoint in both variables and
invariances, quantum gravity in the Ashtekar variables is formulated in
terms of a {\em connection} and {\em all three} constraints are
present. The relation of quantum gravity to Yang-Mills theory is summarized
in the following diagram:
\begin{equation}
\left.
\begin{array}{cl}
\left.
\begin{array}{c}
\hat G \psi = 0
\end{array}
\right]
&
\mbox{YMT}
\\
\left.
\begin{array}{c}
\hat D \psi = 0
\\
\hat H \psi = 0
\end{array}
\right]
&
\mbox{QG}
\end{array}
\right]
\mbox{QG in Ashtekar variables}
\end{equation}

The idea behind the loop representation is to choose a representation of
the quantum algebra of observables on a state space that contains
functionals $\psi[\c]$ of loops $\c$ (as opposed to $\psi[A]$ or
$\psi[g]$).  The point is that the constraints can be treated more easily
in the loop representation than in other representations: (i) The gauge
constraint can be solved by gauge invariant variables (the Wilson loops)
already on the classical level. (ii) The diffeomorphism constraint can be
solved by considering states that are knot invariants, i.e. states that
only depend on the diffeomorphism equivalence class of a loop. And (iii),
the Hamiltonian constraint, also known as the Wheeler-DeWitt equation, has
non-trivial solutions in terms of loops with intersections.  This is, in a
nutshell, the message we want to clarify. The loop representation of
quantum gravity has also very interesting features beyond the constraints,
which we will mention only briefly.

In section 2, we give a naturally very brief history of loops in
mathematics (knot theory) and physics (gauge theory), and we argue why loop
states can be expected to be useful in quantum gravity. In section 3, we
discuss Maxwell theory as an example for a theory that possesses a complete
formulation in the loop representation. In section 4, we introduce the loop
representation for quantum gravity, and discuss various steps of the
program of canonical quantization in the loop representations. In section
5, we conclude with a few comments on loop representations in general and
on the status of the loop representation for general relativity.

%%%%%%%%%%%%%%%%%%%%%%%%%%%%%%%%%%%%%%%%%%%%%%%%%%%%%%%%%%%%%%%%%%
\section{History of Loops}

For the purpose of motivating the loop representation of quantum gravity,
we first introduce loops in the context of knot theory and gauge
theory. A starting point for these two topics can be found in the early
work by Faraday and Gauss.

\begin{figure}
\centerline{\mbox{\epsfig{figure=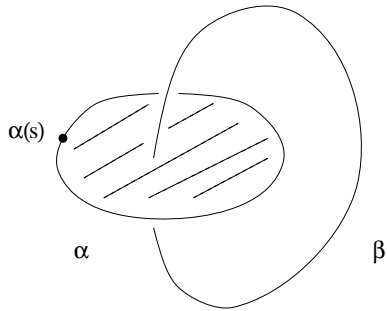,width=50mm}}}
\caption{Two loops $\alpha$ and $\beta$ that link once.}
\label{f0}
\end{figure}

Before entering the discussion, let us define loops.  First, we define a
path $\mu$ as a continuous, piecewise smooth map
$\mu:[s,t]\rightarrow\Sigma$ from an interval into the three-manifold
$\Sigma$. We usually choose $\Sigma = \reals^3$. A loop is a closed path
(figure \ref{f0}), $\a:[0,1]\rightarrow\Sigma$ such that $\a(0)=\a(1)$.
Equivalently, $\a$ is a map from the circle into the three manifold.

We denote the path from
$\a(s)$ to $\a(t)$ along $\a$ by $\a_s^t$. The parametrization implies an
orientation along the loop, e.g. $\a(0)$ lies on $\a_s^t$ only if $s>t$.
The inverse $\a^{-1}$ of a loop $\a$ is defined by reversing the
parametrization,
\beq
	\a^{-1}(s) = \a(1-s).
\eeq
If two loops $\a$ and $\b$ intersect at a point, i.e. $\a(s) = \b(t)$ for
some $s \neq t$, then we can define a combined loop $\c = \a_s\!\circ_t\b$,
as the loop obtained by first going around $\a$ from $\a(s)$ to $\a(s)$ and
then around $\b$ from $\b(t)$ to $\b(t)$. If the point of combination is
clear from the context we may just write $\c = \a\b$. For example, if
$\a(1) = \b(0)$,
\beq
	\c(s) = \left\{
	\begin{array}{ll}
	\a(2s)   &   \mbox{if $0\leq s < 1/2$} \\
        \b(2s-1) &   \mbox{if $1/2 \leq s \leq 1$}
	\end{array}
	\right.
\eeq
where we have made the new parametrization explicit. Paths are combined
analogously, usually at their endpoints. Finally, we also consider
multiloops, which are unordered collections of loops. Given two loops $\a$
and $\b$, we denote the multiloop $\eta:S^1\times S^1\rightarrow\Sigma$
containing $\a$ and $\b$ once by $\eta=\a\cup\b$ (= $\b\cup\a$).

\subsection{Loops in the Work of Faraday and Gauss}

During the years of 1821--32, when M.\ Faraday was working on
electrodynamics, he also developed the concept that the electromagnetic
forces are transmitted through a force field, and that this force field has
physical reality \cite{Be74}. The competing point of view on
electromagnetic forces at that time was that of 'action at a distance', in
particular that no additional physical effect comes between charges and
their relative forces. The idea to use force fields as intermediaries was
not new, but fields were considered to be just a useful mathematical
tool. As it often happens when a mathematical construction captures the
essentials of a physical phenomenon, it becomes part of our intuition about
what is actually physically present. We are justified to think of the
electric field as 'real' because we can, for example, store energy in it or
compute its propagation in electromagnetic waves. (At a deeper level of
'reality' we will have to deal with QED.)

What is relevant here is that Faraday also noted that in the absence of
sources the field lines have to close on themselves, i.e. to form loops,
since only then the electric field is divergence free,
\beq
	D_aE^a(x) = 0.
\eeq
Furthermore, we can argue that the elementary excitation of the electric
field is based on loops. We cannot construct a divergence free vector field
with support on only a point, but it works one dimension up. Of course, we
usually consider electric fields with three-dimensional support, but gauge
invariance requires only loops, not three-spaces.

This simple fact has applications in modern gauge theories through the
so-called Wilson loops, which are gauge invariant variables defined through
the parallel transport of spinors around loops. We will discuss the role
played by loops in gauge theory in section 2.3.

What may be less known is that electromagnetism also lead to the study of
the second type of invariance that we want to consider, namely
diffeomorphism invariance. On January 22, 1833, C.F.\ Gauss found the
answer to the following problem \cite{Ga1833}, which reads like a recent
text book problem, although the implications are quite subtle. What is the
work done on a magnetic pole which is moved on a closed curve in the
presence of a current loop? Notice that there are two loops in the problem,
say $\a$ and $\b$, which we assume to be non-intersecting (see figure
\ref{f0}). The answer can be expressed in terms of what is now known as the
Gauss linking number,
\beq
	gl(\a,\b) = \frac{1}{4\pi} \int ds \int dt \e_{abc}
		\dot\a^a(s) \dot\b^b(t)
		\frac{\a^c(s) - \b^c(t)} {|\a(s) - \b(t)|^3}.
\label{gl}
\eeq
Although not obvious when written this way, $gl(\a,\b)$ is an integer. The
Gauss linking number counts (with an appropriate sign) how often two loops
are linked, or equivalently, how often one loop winds around the other. If
there is no linking, then $gl(\a,\b)=0$.

The key point is that the Gauss linking number does not change under small,
smooth deformations of the loops, that is, the Gauss linking number is
invariant under diffeomorphisms. Gauss himself found it quite remarkable
that an integral as in (\ref{gl}) has this property. The equivalence
classes of loops under diffeomorphism (connected to the identity) is the
topic of a branch of mathematics called knot theory.

Historically, the relation between loops and diffeomorphism invariance was
developed before loops found their way into gauge theories. We therefore
make a few comments on knot theory in section 2.2 and then briefly
discuss loops in gauge theories in section 2.3.

\subsection{Knot Theory}

Knot theory studies the equivalence classes of loops without intersections
under diffeomorphisms that are connected to the identity (e.g.\
\cite{Ka91}). For a single loop the equivalence class is called a knot
class, equivalence classes of multiloops are also called link class.  If
the loops are allowed to intersect, one speaks of generalized knot and link
classes.  We denote knot and link classes by $\{\eta\}$.

\begin{figure}
\par
\centerline{
\epsfig{figure=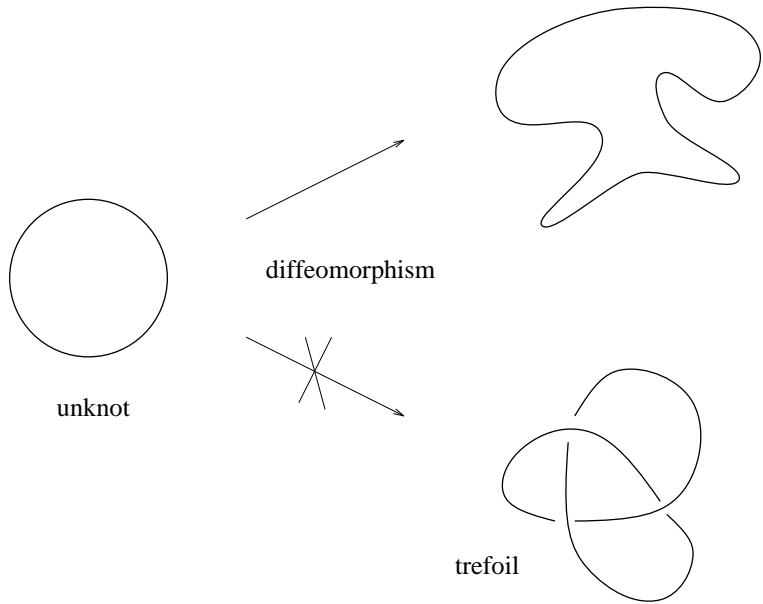,width=100mm}
}
\par
\caption{A diffeomorphism cannot tie a knot into the unknot.}
\label{f1}
\end{figure}

Recall that a diffeomorphism is a $C^\infty$ map between manifolds that is
one-to-one, onto, and has a $C^\infty$ inverse. An example for what a
diffeomorphism can do with the unknot, which is the trivial knot, is shown
in figure \ref{f1}.
A diffeomorphism can deform the unknot quit arbitrarily, but it cannot tie
a trefoil knot in the unknot. The simple intuitive reason is that in a
continuous transformation from one case to the other, two lines would cross,
but a one-to-one and onto map cannot produce intersections or take them
apart.

\begin{figure}
\par
\centerline{
\epsfig{figure=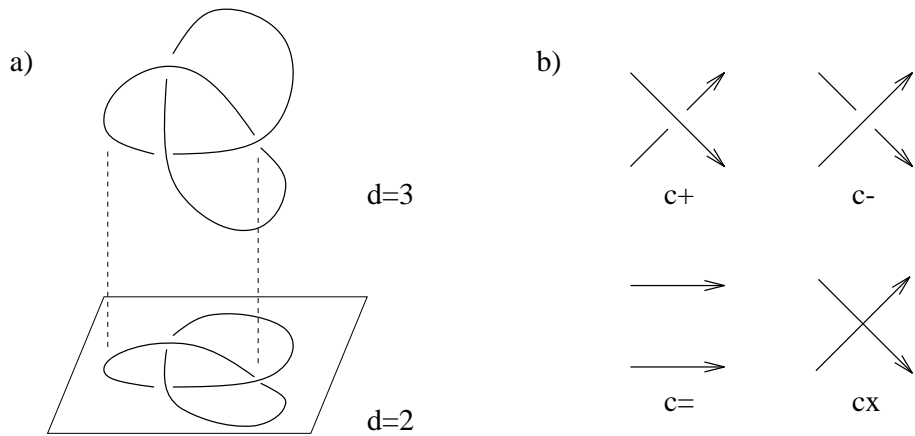,width=120mm}
}
\par
\caption{a) Projecting a knot onto a knot diagram; b) the four crossings in
a knot diagram}
\label{f2}
\end{figure}

One could say that knot theory was founded in 1877 when P.G. Tait
formulated the program of classifying all knots. To this end he introduced
knot diagrams. A knot diagram is a non-degenerate projection of a knot that
lives in three dimensions onto a two dimensional plane plus a prescription
whether an intersection in the plane came from an under crossing or over
crossing of two lines in three dimensions (see figure \ref{f2}).
One can show that there always exists a projection that is non-degenerate
in that no more than two different points are mapped into one. For example,
we are not allowed to project the unknot 'sideways' onto a line segment.

The problem of classifying knots in three dimensions can be shown to be
equivalent to classifying knot diagrams in two dimensions.  An important
tool to establish the equivalence of two knot diagrams are the
Reidermeister moves. As a first step Tait compiled tables of inequivalent
knot diagrams. Figure \ref{tait} shows a copy of Tait's original table from
1884 \cite{Ta1877} for the first seven orders of knottiness. Such tables
are even today an important source for examples.
\begin{figure}
\vspace{17cm}
\caption{``The first seven orders of knottiness'', Tait 1884.}
\label{tait}
\end{figure}

As an aside, the first application of knot classes to physics appeared in
the work of Lord Kelvin in 1869 \cite{Th1869} (which prompted Tait to start
his investigations). He proposed that atoms are 'smoke ring vertices in the
ether'. The stability and variety of atoms are to be directly related to
the equivalence classes of loops under diffeomorphism. The spectral lines
would be created by vibrations of the loop. Today we know about
transmutations of elements, and those could have been discussed in terms of
line crossings (see below).  For about thirty years Kelvin's theory of
'topological matter' was taken seriously, and for example Maxwell concluded
that it accommodated more features of atomic physics than other models.

Returning to the classification problem, another important tool are knot
invariants.  A knot invariant is a functional on the space of loops that
assigns to loops in the same knot class the same number, i.e. $\psi[\eta] =
\psi[\{\eta\}]$.  If $\psi$ is a knot invariant, then
\beq
	\a \sim \b \quad \Rightarrow \quad \psi[\a] = \psi[\b].
\eeq
Therefore, if $\psi[\a] \neq \psi[\b] $ then the loops are not equivalent.
The Gauss linking number is an example for a link invariant.

The hard part is to construct the inverse.  Indeed, one of the central,
unsolved problems of knot theory is to find a {\em complete} set of knot
invariants, $\{\psi_i\}$, such that
\beq
	\psi_i[\a] = \psi_i[\b] \;\forall i
 	\quad \Rightarrow \quad \a \sim \b.
\eeq
However, a complete, indirect classification is possible via the complement
of loops in $\Sigma$. Also, there exist algorithms to generate all knot
classes.

The most important (and most complete) knot invariants arise in the study
of knot polymials.  Knot polynomials were introduced by Alexander in 1928
\cite{Al28}. A knot polynomial $P_q(\c)$ assigns to each knot diagram of a
loop $\c$ a Laurent polynomial in a complex variable $q$ such that
\beqa
	\mbox{(i)}  &&   \mbox{$P_q(\c)$ is a knot invariant},
\\
	\mbox{(ii)} &&  \mbox{$P_q(\mbox{unknot}) = 1$},
\\
	\mbox{(iii)} &&  \mbox{The skein relations are satified.}
\eeqa
In the skein relations, or crossing change formulas, one considers three
knot diagrams that differ only at one crossing. The three different
possibilities are an over crossing, $c_+$, an under crossing, $c_-$, and no
crossing, $c_=$ (see figure \ref{f2}). If intersections are allowed, one
includes the intersections, $c_\times$.  Since a reflection at a plane is a
diffeomorphism which is not connected to the identity, $c_+$ and $c_-$ are
inequivalent.

For example, the Alexander-Conway polynomial $A_q(\c)$ is uniquely
determined by the skein relation
\beq
	A_q(c_+) - A_q(c_-) = q A_q(c_=).
\label{skeinA}
\eeq
One can show that such crossing changes are sufficient to reduce any knot
diagram to the unknot by recursion. As a simple exercise the reader may
show that
\beq
	A_q(\mbox{trefoil}) = 1 + q^2.
\eeq
While the knot polynomial $A_q(\c)$ allows one to distinguish a large
number of knots, it is not complete. For example, the mirror images of
the trefoil lead to the same polynomial.

\subsection{Gauge Theory}

The canonical formulation of non-abelian gauge theories and the application
of Wilson loops is a very interesting topic, and it is reviewed by Loll in
this volume. Here we collect only a minimal set of definitions and facts
that fit our discussion of the loop representation.

The canonically conjugate phase space variables of gauge theory are a
configuration variable, the connection one-form $A_a^i(x)$, and a
momentum variable, the densitized triad $E^{bj}(x)$, on $\Sigma$. Here $a,b,
... = 1,2,3$ are (co-)tangent space indices of $\Sigma$, $i,j,...$ are the
indices of the internal gauge group. For matrices in the Lie algebra of the
gauge group we write for example $A_a = A_a^i \tau^i$, where $\tau^i$ are
the generators of the group (in the fundamental representation). The
Poisson algebra is
\beqa
	\{A_a^i(x),A_b^j(y)\} = 0, \quad && \{E^{ai}(x), E^{bj}(y)\} = 0,
\\
	\{A_a^i(x),E^{bj}(y)\} &=& \d_a^b \d^{ij} \d^3(x,y).
\eeqa

Gauge invariance implies the presence of a constraint in the canonical
formalism, the Gauss constraint
\beq
	G^i(x) \equiv D_a E^{ai}(x) = 0,
\label{gausscon}
\eeq
where $D_a$ is the covariant derivative constructed from $A_a$.

There are different ways to deal with the gauge constraint. For example, we
can choose to perform a gauge-fixing or not, and we can choose to solve the
constraint in the classical theory or in the quantum theory.  Given a gauge
fixing, one has to check whether the final result depends on the gauge or
not, and there may be ambiguities in the quantum theory.  In principle,
these problems do not appear if one can find a gauge invariant formulation,
that is if one is able to solve the constraints classically.

Solving the constraint classically is referred to as reduced phase space
quantization, the alternative is to impose the constraints in the quantum
theory as in Dirac quantization.  In general, the result is {\em not} the
same (e.g.\ \cite{RoTa89}).

The idea leading to loops is to give a reduced phase space formulation of
gauge theories in terms of Wilson loops \cite{Ma62,Wi74,Po79,GaTr80,Mi83}
(and Loll in this volume). A Wilson loop, $h[\c,A]$, is the trace of the
holonomy of $\c$ and $A$,
\beqa
	h[\c,A] &=& \tr U_\c(A),
\\
	U_\c(A) &=& P\exp\int_0^1\!ds\,\dot\c^a(s) A_a(\c(s)),
\eeqa
where $U_\c(A)$ is the matrix for parallel transport of spinors around the
loop. The $P$ denotes path ordering, i.e. for some one parameter family of
matrices $M(s)$,
\beq
       P\exp\int_0^1 ds M(s) = 1 + \int_0^1\!ds\, M(s) + \int_0^1\!ds\,
\int_0^s\!dt\, M(t) M(s) + \ldots,
\label{pexp}
\eeq
such that the products of matrices are always ordered according to the size
of the parameter. (There is no factorial factor.)

The Wilson loops are gauge invariant since under a gauge transformation
$U\rightarrow g U g^{-1}$ and $\tr g U g^{-1} = \tr U$.  This is the reason
why we consider loops and not paths.  What makes the Wilson loops important
is that not only are they gauge invariant, but in a sense they span the
space of all gauge invariant functionals of the connection. There are
reconstruction theorems, for example for $SU(N)$ \cite{Gi81,AsLe93}, of the
type that if the trace of the holonomy $h[\c, A]$ is known for a given
$A$ and for all $\c$, then $A$ is determined up to gauge.

Starting point for the loop representations is that instead of elementary
variables $A_a^i$ and $E^{bj}$ we can choose the loop variables
\beqa
	T^0[\c] &=& \tr U_\c(A),
\label{lova0}
\\
	T^1[\c]^a (s) &=& \tr U_{\c_s^s}(A) E^a(\c(s)),
\label{lova1}
\eeqa
where the $T^1$ variables are obtained by inserting the matrix $E^a$ at the
parameter $s$ into the parallel transport around $\c$ (see also section
4.2). Classically, the choice of variables is equivalent modulo gauge for
$SU(N)$. The question is whether loop variables offer any advantages over
the conventional approaches in the quantum theory.

We could now give the definition of a loop representation for a gauge
theory, but since our main objective is the loop representation of quantum
gravity, let us first complete the picture of how quantum gravity relates
to knot theory and gauge theory.

\subsection{Quantum Gravity}

\subsubsection{Before 1984.}

Let us summarize the situation in knot theory, gauge theory, and general
relativity up to the year 1984 from the perspective of what is important
for the loop representation of quantum gravity:
\begin{itemize}

\item Knot theory: \\
The characteristic invariance is diffeomorphism invariance. The project is
to classify all knots. The status is that generating and labeling all
knots is well understood, but a deep insight into what a complete set of
knot invariants could be is missing.

\item Gauge theory: \\
The characteristic invariance is gauge invariance. The project is to
perform a reduced phase space quantization. The status is that for
non-abelian gauge theories it is not known how to treat the equations of
motions, and around 1980 most people in the field give up on using loop
variables (with the notable exception of the group around Gambini,
see references).

\item General relativity: \\
The characteristic invariance is space-time diffeomorphism invariance,
represented in the canonical formalism by the diffeomorphism and
Hamiltonian constraints. The project is canonical quantization. The status
is that canonical quantization is incomplete, in particular because
one does not know how solve the constraints in the metric variables.

\end{itemize}
Notice that at this time the three topics are completely unrelated. While
diffeomorphism invariance plays a central role in both knot theory and
general relativity, knots play no role in general relativity. And general
relativity is not a gauge theory in the sense explained in section 2.3.

What we are leading up to is, of course, that in the following years a
fruitful combination of all these ideas became possible.  While the
problems mentioned above under status have not been solved, there has been
some progress.

\subsubsection{After 1984.}

Let us now summarize some of the recent developments:
\begin{itemize}

\item
Jones 1985 \cite{Jo85} \\
After Alexander, no new knot polynomials had been found until the discovery
by Jones of the Jones polynomial, $J_q(\c)$. The important point is that his
techniques allow a systematic investigation into the space of knot
invariants. Roughly speaking, each representation of the braid group that
carries a Markov trace defines a knot invariant.

\item
Ashtekar 1986 \cite{As86,As87,As91} \\ Via a canonical transformation from
the metric variables, Ashtekar constructs a new set of canonical
variables. These are an $SL(2,\complexes)$ connection $A_a^i$ and a
conjugate momentum $E^{bj}$, the same type of variables as in a non-abelian
gauge theory. There are as before a spatial diffeomorphism constraint $D$
and a Hamiltonian constraint $H$, and in addition there is a Gauss
constraint $G$ (\ref{gausscon}) (since $g^{ab} = E^{ai} E^{bj}$, which is
invariant under internal gauge transformations).

One advantage of the Ashtekar variables is that the constraints are
considerably simpler as in the metric variables (at least as far as solving
the constraints is concerned, see below).  The price to be paid is that
there is an additional constraint, and that the Ashtekar connection is
complex, and suitable reality conditions have to be imposed to recover real
general relativity.

The perhaps most important feature of the new variables
is that in the Ashtekar formalism the kinematics of general relativity
is imbedded in that of a non-abelian gauge theory for the Ashtekar
connection. We have a standard gauge constraint plus the diffeomorphism and
Hamiltonian constraint. This imbedding allows us to import ideas from
Yang-Mills theory into general relativity, the loop representation being
the prime example.

\item
Rovelli and Smolin 1988 \cite{RoSm88,RoSm90} \\ Rovelli and Smolin took the
idea to use Wilson loops as gauge invariant variables one step further and
constructed a representation of the operator algebra of quantum gravity on
wavefunctions that are functionals of loops, $\psi[\eta]$.  After all that
has been said, it should be obvious that loops have a natural role to play
in canonical quantum gravity. Let us consider the constraints:
\begin{enumerate}
\item We can solve the gauge constraint,
\beq
	G = 0,
\eeq
on the classical level as in the reduced phase space formulation of gauge
theory by using the loop variables $T^0$ and $T^1$, (\ref{lova0},
\ref{lova1}), for the Ashtekar connection.

\item We can solve the diffeomorphism constraint of the quantum theory,
\beq
	\hat D \psi[\eta] = 0,
\label{D}
\eeq
by imposing that $\psi[\eta]$ is a knot invariant. Indeed, since a finite
diffeomorphism $f$ has a natural representation on the support of the
wavefunctions, $(f\cdot\psi)[\eta] = \psi[f^{-1}\circ\eta]$, equation
(\ref{D}) for the generator of diffeomorphisms $\hat D$ implies that
$\psi[\eta]$ must be a knot invariant, $\psi[\eta] = \psi[\{\eta\}]$.

This formal solution of the constraint has to be compared with the metric
representation. In that case one can impose that the wavefunctions are
functionals of equivalence classes $\{g\}$ of metrics under diffeomorphisms
(called three-geometries), $\psi[g] = \psi[\{g\}]$.  However, one of the
major problems with canonical quantization based on the metric variables is
the complicated structure of the space of three geometries. In the case of
the loop representation, the situation is simpler since the knot
invariants can be easily labeled. A similar simplification occurs for the
hydrogen atom, where the energy angular-momentum basis $\psi(n,l,m)$ has
many advantages over the position basis $\psi(x)$.

\item
A priori it is not clear at all whether the Hamiltonian
constraint,
\beq
	\hat H \psi[\eta] = 0,
\label{H}
\eeq
can be solved in the loop representation. One reason why the metric
representation is incomplete is precisely that we do not know how to
factor-order, regularize and solve the Wheeler-DeWitt equation. The
encouraging fact in the loop representation is that there exist at least
some trivial solutions to the Wheeler-DeWitt equation, while for the metric
variables not a single solution to the full equation had been known.
As noted by Rovelli and Smolin based on (\cite{JaSm88}), the Hamiltonian
constraint trivially annihilates all loop functionals that have support
only on non-intersecting loops.

\end{enumerate}

At this point,
the loop representation seems to be a promising approach to canonical
quantum gravity. The Gauss and diffeomorphism constraints can be solved
formally by loop techniques, and the Wheeler-DeWitt equation is at least
not completely incompatible with loop functionals.

\item Witten 1989 \cite{Wi89} \\
Witten showed that the Jones polynomial, which is tied to two dimensions
through the knot diagrams, can be computed as the vacuum expectation value
of the Wilson loops in Chern-Simons field theory, which is a topological
field theory in three dimensions:
\beq
	J_q(\c) \sim \langle h[\c,A_{CS}] \rangle_{CS},
\eeq
where $A_{CS}$ is the Chern-Simons connection. There are many important
aspects to this work, here we want to focus on the new link between
invariants in two and three dimensions.
While Witten's proof is abstract in the sense that he does not directly
evaluate the path integral that defines the expectation value (and
mathematicians are still struggling to make his arguments rigorous), one
can show that in a perturbation theory based on the path integral, the
leading terms satisfy the skein relation that define the Jones polynomial
\cite{Sm89,GuMaMi90}.
In this expansion one obtains knot invariants in analytic
form analogously to the integral (\ref{gl}) for the Gauss linking number,
as opposed to the recursion formulas for knot diagrams.

\end{itemize}
The above results are very interesting and far reaching for many more
reasons than just the canonical quantization of gravity. One of the reasons
why the author finds the loop representation appealing is that we can tie
together all we have said about knot theory and the loop representation,
and actually use knot theory (beyond formal solution of the diffeomorphism
constraint) to construct non-trivial solutions to the Wheeler-DeWitt
equation:
\bi
\item Br\"ugmann, Gambini and Pullin 1992
\cite{BrGaPu92a,BrGaPu92b,BrGaPu93} \\
One can generalize the Jones polynomial to the case of non-intersecting
loops. The action of the Hamiltonian constraint on the analytic knot
invariants arising in the perturbation theory of Chern-Simons theory can be
computed explicitly. In fact, the second coefficient $a_2(\c)$ of the
Alexander-Conway polynomial is annihilated by the Hamiltonian constraint.
As explained in section 4.6, the non-triviality of this solution lies in
the fact that because of the use of intersecting loops it is not
simultaneously annihilated by the determinant of the metric.

Further understanding of the structure of the space of solutions can be
gained by considering a transformation between the connection
representation and the loop representation. If there exists a suitable
transform, then the Jones polynomial (and not just a few coefficients) is a
solution to the Hamiltonian constraint with a non-vanishing cosmological
constant.  This one solution for a cosmological constant leads to several
solutions for vanishing cosmological constant.
\ei
This concludes the motivational part why the loop representation is natural
in quantum gravity. In the remainder of this lecture we discuss some
details of the loop representation in the context of an example, Maxwell
theory, and for quantum gravity itself.

%%%%%%%%%%%%%%%%%%%%%%%%%%%%%%%%%%%%%%%%%%%%%%%%%%%%%%%%%%%%%%%%%%
\section{Example for Complete Loop Representation: Maxwell Theory}

The loop representation of Maxwell theory is an ideal example for an
introduction to loop representations. First of all, in this simple case
the program of canonical quantization can be rigorously completed in the
loop representation. Furthermore, the standard formulation in terms of Fock
space is well understood, and we can compare the two representations and
obtain a physical interpretation for the loop operators. In fact, the two
representations turn out to be equivalent, since there exists a faithful
transform between them.

The earliest work on a loop variable formulation for a $U(1)$ gauge theory
is that of Mandelstam \cite{Ma62}, and by Gambini and Trias
\cite{GaTr80,GaTr83}. We will follow closely Ashtekar and Rovelli
\cite{AsRo92}, where the loop representation is constructed in a form and
with techniques that are directly related to the loop representation of
quantum gravity.

\subsection{Bargmann Representation of Maxwell Theory}

We consider a pure $U(1)$ gauge theory on $\reals^3$ with a flat background
metric.  The canonical variables defining the phase space are a one-form
$A_a(x)$ and an electric field $E^a(x)$ such that $\{A_a(x), E^b(y)\} =
\d_a^b
\d^3(x,y)$. There is one first class constraint,
\beq
	D_aE^a(x) = 0.
\eeq
By fixing the gauge, $D^aA_a(x)=0$, we pass to the reduced phase space
which is parametrized by divergence free fields $A_a^T(x)$ and $E_a^T(x)$.

Such transverse fields are conveniently described in the
momentum representation. Let us introduce canonical variables $q_j(k)$ and
$p_j(k)$, $j=1,2$, on momentum space with the only non-vanishing Poisson
bracket $\{q_m(-k),p_n(k')\} = \d_{mn}\d^3(k,k') $. In terms of
complex polarization vectors $m_a(k)$, $m_ak^a=m_am^a=0$ and $m_a\bar m^a =
1$, we can write the transverse variables as
\beqa
	A_a^T(x) &=& \frac{1}{(2\pi)^{3/2}} \int d^3 k e^{ik\cdot x} (q_1(k)
m_a(k) + q_2(k) \bar m_a(k)),
\\
	E_a^T(x) &=& - \frac{1}{(2\pi)^{3/2}} \int d^3 k e^{ik\cdot x} (p_1(k)
m_a(k) + p_2(k) \bar m_a(k)).
\eeqa

In order to make a direct transition from the Fock representation to the
loop representation possible, we have to work with Bargmann coordinates,
$\z$ and $\bar\z$, defined by
\beq
	\z_j(k) = \frac{1}{\sqrt 2} (|k| q_j(k) - ip_j(k)).
\eeq
The non-vanishing Poisson bracket is now
\beq
	\{ \z_m(k), \bar\z_n(k') \} = i|k| \d_{mn} \d^3(k,k').
\eeq
The one dimensional analog of these variables are the Bargmann variables
$z$ and $\bar z$ for the simple harmonic oscillator, $z = \frac{1}{\sqrt 2}
(\omega q - ip)$. The Bargmann variables $\z$ and $\bar\z$ are directly
related to the positive frequency part of the transverse fields $A_a^T$ and
the negative frequency part of $E_a^T$,
\beqa
	\mbox{ }^+ A_a^T(x) &=& \frac{1}{(2\pi)^{3/2}} \int \frac{d^3 k}{|k|}
e^{ik\cdot x} (\z_1(k) m_a(k) + \z_2(k) \bar m_a(k)),
\\
	\mbox{ }^- E_a^T(x) &=& - \frac{i}{(2\pi)^{3/2}}
\int \frac{d^3 k}{|k|}
e^{ik\cdot x} (\bar\z_1(-k) m_a(k) + \bar\z_2(-k) \bar m_a(k)).
\eeqa

The quantum theory can be formulated in the Bargmann representation, where
states are holomorphic functionals $\Psi[\z] \equiv \Psi[\z_1, \z_2]$ of
the Bargmann variables. We represent the operators corresponding to the
Bargmann variables via
\beqa
	\hat\z_j(k) \Psi[\z] & = & \z_j(k) \Psi[\z] ,
\\
 	\hat{\bar\z}_j(k) \Psi[\z] & = &
	-i\hbar |k| \frac{\d}{\d\z_j(k)} \Psi[\z].
\eeqa
which implies that, as required, the canonical commutation relations
are satisfied. Again, compare with the harmonic oscillator where $
\hat z \Psi(z) = z \Psi(z)$ and
$ \hat{\bar z} \Psi(z) = - i \hbar \omega (d/dz) \Psi(z) $.

To complete the mathematical setup, we have to specify an inner product
that turns the space of states into an Hilbert space. This poses in general
a non-trivial problem, see below. Here, an inner product exists and is
uniquely determined by the reality conditions $\hat{\bar\z}_j(k) =
\hat\z^*_j(k)$. This inner product is the unique, Poincar\'e invariant
inner product of the Bargmann representation,
\beqa
	\langle \Phi[\z] | \Psi[\z] \rangle
	& = &
	\int d\mu(\z,\bar\z) \overline{\Phi[\z]} \Psi[\z],
\\
	d\mu(\z,\bar\z)
	&=&
 	\prod_j d\! I \z_j(k) \wedge d\! I \bar \z_j (k)
	\exp \left( - \frac{1}{2\hbar} \int \frac{d^3 k}{|k|}
	|\z_j(k)|^2 \right) .
\label{measure}
\eeqa
For the harmonic oscillator, the measure is of the type
$ dz \wedge d\bar z e^{-z\bar z}$.

The Bargmann representation has the usual interpretation in terms of Fock
states. The vacuum $|0\rangle$ and the one photon state $|k,\e\rangle$ for
momentum $k$ and helicity $\e$ are given via $\Psi[\z] = \langle \z | \Psi
\rangle$ by
\beqa
	|0\rangle: && \quad \Psi_0 [\z] = 1 ,
\\
	|k,\e\rangle: && \quad \Psi_{k,\e} [\z] = \z_\e(k) .
\eeqa
The multiplication operator $\hat\z_j(k)$ acts as creation operator in the
Fock basis, the derivative operator $\hat{\bar\z}_j(k)$ as annihilation
operator. For example, the one photon state is created by action with
$\hat\z_j(k)$ on the vacuum, $\Psi_{k,\e} [\z] \equiv \hat\z_j(k)
\Psi_0[\z]$.

\subsection{Loop Representation of Maxwell Theory}

Let us consider the same $U(1)$ gauge theory as in the preceding section, but
now we choose loop variables instead of $\z$ and $\bar \z$.
Because the
gauge group is abelian, the loop variable $T^0$ (\ref{lova0}) simplifies,
and a simpler choice than $T^1$ (\ref{lova1}) is possible.
We define
\beqa
	h[\a] &=& \exp \oint_\a\!ds^a\: \mbox{}^-\!A_a^T,
\\
	E[f] &=& \int_{{\reals}^3}\!d^3x\, f_a\, \mbox{}^+\!E^{Ta},
\eeqa
where $\a$ is a loop and $f_a$ a one-form.  The loop variable $h[\a]$ is
the abelian holonomy and takes the place of $T^0[\a]$. The analog of
$T^a[\a](s) = \tr U_\a E^a(\a(s))$ is $h[\a] \mbox{}^+\!E^a(\a(s))$, but
since for $U(1)$ gauge invariance no trace is necessary, we can use the
loop-independent (smeared) variable $E[f]$. (Note that $\tr E^a = 0$ for
$SU(N)$.)

Our choice of loop variables high-lights the fact that there are different
possibilities to construct loop variables. In fact, we could as well start
with $h[\a]$ and $h[\a] \mbox{}^+\!E[f]$ as our elementary variables. Of
course, the interpretation of the momentum operators is then different, and
not as nicely related to the Fock representation. The choice of the
negative frequency connection for $h[\a]$ and the positive frequency
electric field is necessary, since other frequency splittings are not
consistent. Still another choice is to use anti-self-dual connections and
self-dual electric fields, as we actually do in gravity.

Notice that the loop variables $h[\a]$ and $E[f]$ are in two ways
overcomplete. First of all, not each label $\a$ and $f$ corresponds to
different variables. Two different loops $\a$ and $\b$ may lead to $h[\a] =
h[\b]\;\forall A$, e.g. if $\b = \a \eta\eta^{-1}$ for some path $\eta$.
And $f_a$ and $f_a + \partial_ag$ give the same $E[f]$. In addition, there is
the non-linear identity
\beq
	h[\a] h[\b] = h[\a\eta\b\eta^{-1}]
\label{spinidu1}
\eeq
for any path $\eta$ that connects $\a$ and $\b$; we also write $h[\a]h[\b]
= h[\a\#\b]$.  The latter type of identities play an important role
since they differ for different gauge groups (compare (\ref{spinidsu2}) for
$SL(2,\complexes)$).

The Poisson bracket algebra is (using the same simplectic structure as
before)
\beqa
 &&	\{h[\a],h[\b]\} =0, \quad \{E[f], E[g]\} = 0,
\label{al1}
\\
 && 	\{h[\a], E[f] \} = ( \oint_\a ds^a f_a ) h[\a].
\label{al2}
\eeqa

Quantization in the loop representation is based on a space of states
$\psi[\a]$ which are functionals of loops. How can we represent the
operators $\hat h[\a]$ and $\hat E[f]$? Since the space of loops is {\em
not} identical with the classical configuration space, which is the space
of connections, the representation will not be the 'usual' one in terms of
multiplication and derivative operators. Perhaps one should ask the
question of how to find a representation with a different emphasis: Is
there any chance at all that there exists a sensible representation for our
unconventional choice of elementary variables?

Rovelli and Smolin suggested in their work on quantum gravity \cite{RoSm90}
that a transform between connection functionals and loop functionals can be
used to derive the loop representation from the connection representation.
An analog in quantum mechanics is the Fourier transform between the
position and the momentum representation, which allows us to transfer both
states and operators, e.g.
\beqa
	\psi(k) &=& \int dx e^{ikx} \psi(x),
\\
	ik & \leftrightarrow & \frac{d}{dx}.
\eeqa
The Rovelli-Smolin loop transform is
\beq
	\psi[\c] = \int d\mu(A) h[\c,A] \psi[A].
\label{lt}
\eeq
Given a functional $\psi[A]$ in the connection representation, the
integration over all connections (modulo gauge) leaves only the loop
dependence in $h[\c,A]$, and the result is a loop functional. The problem
in general, and in particular for the $SL(2,\complexes)$ connections of
general relativity, is that {\em the measure $d\mu(A)$ is not
known}. Without a definition of the measure, the loop transform is
completely formal.

For Maxwell theory, however, we do have a measure explicitly available. We
therefore can compute the transform from states in the Bargmann
representation to the loop representation via
\beq
	\psi[\c] = \int d\mu(\z,\bar\z) h[\c] \Psi[\z],
\eeq
where the measure is defined in (\ref{measure}).  An explicit calculation
leads to the following result for the transform of a Bargmann state, and
this result is of central importance for the relation between the two
representations. Ashtekar and Rovelli show that
\beq
	\psi[\c] = \Psi[F_j(\c,k)],
\label{psiPsi}
\eeq
where $F_j(\c,k)$ is obtained from the Fourier transform of
the so-called formfactor $F^a(\c,x)$,
\beqa
	F^a(\c,x) &=& \int ds \dot\c^a(s) \d^3(x,\c(s)),
\\
	& = &  \frac{1}{2\hbar(2\pi)^{3/2}} \int d^3k e^{ik\cdot x}
	(F_1(\c,k) m^a(k) + F_2(\c,k) \bar m^a(k)).
\eeqa
In words, given a loop $\c$ we compute the number $\psi[\c]$ for a state in
the loop representation that was obtained via the transform of a state
$\Psi[\z]$ in the Bargmann representation by evaluating $\Psi[\z]$ for $\z
= F_j(\c,k)$.

The name 'form factor' is appropriate since according to
\beq
	\oint_\c ds^a f_a = \int ds \dot\c^a(s) f_a(\c(s)) =
	\int d^3x F^a(\c,x) f_a(x),
\label{intint}
\eeq
the factor $F^a(\c,x)$ allows us to separate the dependence on the loop
from the dependence on the one-form.  In the loop transform, the loop is
combined with the connection (expressed in terms of the Bargmann
coordinates) in precisely the form (\ref{intint}), and after the
integration over one-forms in the loop transform, only the loop dependence
remains. Therefore it is plausible that, as in (\ref{psiPsi}), the
transform of an arbitrary state $\Psi[\z]$ depends on the loop only through
the form factor.

Since we can explicitly compute the transform of any state, we can define
an operator $\hat O_L$ in the loop representation by the transform of $\hat
O_B \Psi[\z]$ in the Bargmann representation. For some operators, however,
we do not have to be able to perform an explicit calculation. In fact,
\beqa
	\hat h_L[\a] \psi[\c] &:=& \int d\mu h[\c] (\hat h_B[\a] \Psi) [\z]
\\
	&=& \int d\mu (\hat h_B^*[\a] h)[\c] \Psi[\z]
\\
	&=& \int d\mu h[\a]h[\c] \Psi[\z]
\\
	&=& \int d\mu h[\a\#\c] \Psi[\z]
\\
	&=& \psi[\a\#\c].
\eeqa
All we have to know in this calculation is that $\hat O_B$ is self-adjoint
with respect to the measure, and we have to know an operator $\hat O_L$
such that
\beq
	\hat O_L h[\a] = \hat O_B h[\a].
\label{transfer}
\eeq

The resulting representation is defined by
\beqa
	\hat h[\a] \psi[\c] &=& \psi[\a\#\c] ,
\label{re1}
\\
	\hat E[f] \psi[\c] &=& \left( i\hbar \oint_\c ds^af_a \right)
	\psi[\c] .
\label{re2}
\eeqa
The $\hat h[\a]$ operator acts by adding the loop $\a$ to the argument of
the loop state, and the $\hat E[f]$ acts by multiplying the loop state with
the loop integral of $f_a$.

Several remarks on the construction of the loop representation via the
Rovelli-Smolin loop transform are in order.
\begin{enumerate}

\item Ashtekar and Rovelli prove that the transform is faithful. Since no
information is lost in the transition from the Bargmann representation to
the loop representation, we have that {\em the loop representation is
equivalent to the Fock representation}.

\item It is easy to check that we have correctly represented the classical
Poisson algebra (\ref{al1},\ref{al2}), in particular that $ [\hat
h[\a],\hat E[f]] =i\hbar \widehat{\{h[\a],E[f]\}} $.  We can therefore
define the loop representation {\em without} ever introducing the Bargmann
representation.

\item The loop representation for quantum gravity (but also for
other gauge groups) can be constructed by {\em assuming} that the loop
operators are self-adjoint, since for the loop operators we do know
transfer relations (\ref{transfer}) \cite{RoSm90}. Furthermore, one finds
that the operators so obtained have the correct commutator
algebra. Consequently, while the loop representation can be motivated by a
formal transform, the definition of the loop representation is independent
of that transform.

\end{enumerate}
Since for Maxwell theory we know a faithful transform, we can answer the
question what the translation of the physical interpretation of the theory
in terms of photons is in the loop representation. We find with
$\psi[\c] = \langle \c | \psi \rangle$ that
\beqa
	|0\rangle: && \quad \psi_0 [\c] = 1 ,
\\
	|k,\e\rangle: && \quad \psi_{k,\e} [\c] = F_\e(\c,k) .
\eeqa
Hence we have discovered a formulation of Maxwell theory in which the
elementary quantum excitations of the electric field are based on loops,
just as we have argued for Faraday's picture of closed field lines. Is this
result helpful for our intuition about physics? Even though the two
representations are equivalent, and the loop representation for Maxwell
theory is structurally appealing, the photon picture is much closer to how
we think about experiments.  However, for a non-abelian gauge theory and in
the absence of a background metric (so that for example the split into
positive and negative frequency parts is not possible), there does not
exist a Fock representation. Nevertheless, it is quite possible that a loop
representation still exists, and in this case, the loop picture may become
a part of our intuition (see section 5.2).

%%%%%%%%%%%%%%%%%%%%%%%%%%%%%%%%%%%%%%%%%%%%%%%%%%%%%%%%%%%%%%%%%%
\section{Loop Representation of Canonical Quantum Gravity}

\subsection{Algebraic Quantization}

Before we introduce the loop representation for canonical quantum gravity,
it is appropriate to recall the program of algebraic quantization and point
out the choices that lead to the loop representation. Our discussion of
Maxwell theory already serves as an example for algebraic quantization, but
the framework described below is designed with the kind of generality
appropriate for general relativity.

The program of canonical quantization of constraint systems, in its
algebraic form due to Dirac \cite{Di65} and Ashtekar \cite{As91},
can be summarized as follows. Given is the classical phase space with the
Poisson algebra of classical observables, which are functions from the
phase space into the complexes, and constraints $C$.
\begin{description}

\item[$\rightarrow$]
Choose a set $S$ of elementary variables which is complete in that it
coordinatizes the phase space, and which is closed under the Poisson
bracket.
\bi
\item[$\bullet$]
These variables will become the elementary operators of the quantum theory.
For a particle on the line we can choose $S=\{1,q,p\}$.  On the one hand,
$S$ has to be large enough such that the elements of $S$ coordinatize the
phase space. On the other hand, $S$ has to be small enough so that we are
able to consistently impose the non-commutative structure of the quantum
theory. In particular, if $S$ consists of all polynomials in the canonical
variables, then there does not exist an irreducible representation of the
quantum algebra (van Hove theorem, \cite{Ho51}).
\item[$\bullet$]
We do not have to choose the canonical variables, for instance in
gauge theory we can choose the loop variables instead.
\ei

\item[$\rightarrow$]
Elevate the elementary variables to operators of the quantum theory, i.e.
form the free algebra $\cal A$ of elements of $S$, $f\mapsto\hat f
\: \forall f\in S $, such that
\beqa
	[\hat f, \hat g] &=& i\hbar \widehat{\{f,g\}} \quad\forall f,g\in S,
\label{ccr}
\\
	\hat f \hat g + \hat g \hat f &=& 2 \widehat{fg}
	\quad\forall f, g:\, fg \in S.
\label{antic}
\eeqa
\bi
\item[$\bullet$]
The relation (\ref{ccr}) is the reason why we call the procedure
'quantization'. The classical Poisson algebra of commuting observables
determines a non-commutative algebra of quantum operators.
\item[$\bullet$]
Anti-commutation relations like (\ref{antic}) arise if the
set $S$ is overcomplete, since if $fg\in S$ we need a unique definition
of $fg \mapsto \widehat{fg}$. More general relations are possible.
The loop variables are overcomplete in that way, compare
(\ref{spinidu1},\ref{spinidsu2}). For a particle on the circle,
e.g.\ $S= \{ 1, \cos\phi, \sin\phi, p \}$, $\cos^2\phi + \sin^2\phi = 1$.
\ei

\item[$\rightarrow$]
Choose a linear representation of $\cal A$ on a vector space $V$.
\bi
\item[$\bullet$]
The vector space is called the space of states, $V=\{\psi\}$.  Again, we
have a choice, and for the loop representation we choose
$V=\{\psi[\eta]\}$, and we choose a particular representation, e.g.\ as for
Maxwell theory in (\ref{re1},\ref{re2}). Notice that in the
infinite-dimensional case we do not have a Stone-von-Neumann theorem
guaranteeing unitary equivalence of representations. Hence, a different
choice of elementary variables and representation can in general lead to
{\em inequivalent} quantum theories.  In other words, different choices can
describe different physical systems, e.g. free electrons and electrons in a
superconductor. In the case of the loop representation there are
indications that the loop transform may relate the different
representations to each other.
\ei

\item[$\rightarrow$]
Represent the constraints,
\beq
	C \mapsto \hat C.
\eeq
\bi
\item[$\bullet$]
While the preceding steps can be performed for a judicious choice of
variables, typically it is in the representation of the constraints that
serious problems arise. First of all, it may not be trivial to express the
constraints in terms of the elementary variables, say in terms of loop
variables. Second, we have to define a factor ordering and usually also a
regularization.
\ei

\item[$\rightarrow$]
Solve the constraints, i.e. find the space of solutions, $V_{sol}\subset
V$, on which
\beq
	\hat C \psi = 0.
\eeq
\bi
\item[$\bullet$]
The space of solutions is sometimes called the space of physical
states, but we prefer to reserve the predicate 'physical' to states in the
Hilbert space of solutions. Canonical quantization calls for the
construction of a Hilbert space of states, and states in $V_{sol}$ may not
be normalizable.
\ei

\item[$\rightarrow$]
Define the Hilbert space ${\cal H}_{phys}$ of physical states. To this end we
have to find an inner product on a subset of $V_{sol}$ such that suitable
$*$-relations on $\cal A$ are satisfied.
\bi
\item[$\bullet$]
For quantum gravity we do not have a criterion like Poincar\'e invariance
that allows us to pick out an unique inner product. There is a conjecture
that if we know a complete set of $*$-relations, for example, if we know
which operators are supposed to be self-adjoint (i.e. observables), then
the inner product is determined uniquely. Such a criterion suffices in most
examples, Maxwell theory included. Its merit is background independence,
and therefore it can be applied in quantum gravity. There the main problem
is existence rather then uniqueness \cite{Re93}.
\item[$\bullet$]
One often introduces an inner product at an earlier stage in the
quantization program before the constraints have been solved (see Hajicek
on pre-quantization in this volume). This may be important for the
regularization of the constraint operators and the solution of the
constraints. However, such an inner product is in general not related to
the inner product of ${\cal H}_{phys}$.
\ei

\item[$\rightarrow$]
Define observables of the quantum theory, define the measurement
process, give an interpretation, make predictions.
\bi
\item[$\bullet$]
The previous step has completed the mathematical setup of the theory. Note
that for quantum gravity, not even the mathematical part has been completed
(the analog is true for other approaches to quantum gravity).  This step
summarizes all the really important, physical questions, and it is kept so
short only because in the case of quantum gravity so little is known about
it.
\ei
\end{description}
Now that a concrete proposal for a quantization program has been put on the
table, let us quickly point out what such a program cannot be. As a matter
of principle, there cannot exist a program into which we feed a classical
theory and by turning a crank or letting a computer run we produce the
corresponding, unique quantum theory. The reason is that quantum theory is
the fundamental theory of which the classical world is a special limit. One
and the same classical theory may be obtained as the limit of inequivalent
quantum theories. Only if we specify information {\em external} to the
classical theory, then we can hope to find a unique quantum theory.

A quantization program is similar to a computer program that requires
additional input at various places in order to run at all. Relevant input
for the loop representations is the choice of loop variables and the choice
of loop states. As already mentioned, different choices may lead to
different quantum theories, and only in experiments can we find out whether
our theory describes the model that we had in mind.

In this sense the program of algebraic quantization of general relativity
is more like a recipe that is justified only if it is successful. If the
steps of quantization listed above can be taken, even excluding the
'physical' one, then this would constitute a major success. If algebraic
quantization in this form fails, then we have {\em not}, of course, shown
that quantization of general relativity is not possible.

Notice that we have resisted the temptation to number the steps. An obvious
possibility is that they may have to be arranged in a different order, or
more seriously that they cannot be separated as shown. For example, we may
need an inner product to make the discussion of the constraints
rigorous. Or we may have to solve the conceptual, physical problems like
the meaning of time in quantum gravity (see Isham in this volume) before we
can even formulate a quantization procedure.

After these remarks of caution about the nature of the proposed
quantization procedure, let us list the steps that we will discuss for
quantum gravity in the loop representation in the next sections. We choose
loop variables as our elementary variables, and we can form the quantum
algebra $\cal A$ (section 4.2). We define a representation of the loop
operators on the space of loop functionals such that the relations
(\ref{ccr},\ref{antic}) for the commutators and Poisson brackets are
satisfied (section 4.2). We define the constraints in the loop variables
(section 4.3) and express the constraint operators as differential
operators on loop functionals (section 4.4). We explain how certain
analytical knot invariants (section 4.5) can give rise to solutions to the
constraints (section 4.6). Finally, in section 5 we give a critical
appraisal of the status of the loop representation.

\subsection{Loop Variables}

\begin{figure}
\par
\centerline{
\epsfig{figure=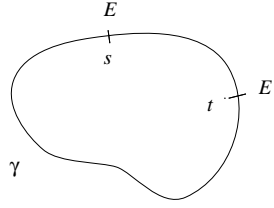,width=35mm}
}
\par
\caption{Insertion of $E$ at parameters $s$ and $t$ along a loop $\gamma$.}
\label{f3}
\end{figure}

The Rovelli-Smolin loop variables for general relativity are
(\cite{RoSm90,As91}, compare section 2.3)
\beqa
	T[\c] & = & \tr U_\c,
\\
	T^a[\c](s) & = & \tr U_{\c_s^s} E^a(\c(s)),
\\
	T^{ab}[\c](s,t) &=& \tr U_{\c_t^s} E^a(\c(s)) U_{\c_s^t}
	E^b(\c(t)),
\\
	& \vdots &
\nonumber
\eeqa
where $U_\c$ is the parallel transport for the $SL(2,\complexes)$ Ashtekar
connection and $E^a$ its conjugate momentum. Each $T^n$ variable is
characterized by $n$ insertions of the momenta into the trace of the
holonomy (see figure
\ref{f3}).

The loop variables are overcomplete. Since for any $SL(2,\complexes)$
matrices $A$ and $B$, $\tr A \tr B = \tr AB + \tr AB^{-1}$, we have for
example that
\beq
	T[\a] T[\b] = T[\a\b] + T[\a\b^{-1}],
\label{spinidsu2}
\eeq
if $\a$ and $\b$ intersect so that $\a\b$ exists. This identity is the
Mandelstam identity (also called spinor identity) for $SL(2,\complexes)$,
compare (\ref{spinidu1}) for $U(1)$. We will incorporate the spinor
identity into the quantum theory via the anti-commutation relations, but in
principle one can remove the overcompleteness by a different choice of
variables \cite{Lo91}.

The Poisson algebra has the structure
\beq
	\{ T^m, T^n \} \sim T^{m+n-1},
\eeq
where '$\sim$' means that the bracket can be expressed in terms of linear
combinations and integrals of loop variables of the given order. The
algebra of the $T^0$ and $T^1$ closes:
\beqa
	\{ T[\a], T[\b] \} &=& 0,
\\
	\{ T^a[\a](s), T[\b] \} &=& \int du \dot\b^a(u) \d^3(\a(s),\b(u))
\nonumber
\\
	&& \quad
	(T[\a_s\!\circ_u\b] - T[\a_s\!\circ_{1-u}\b^{-1}]),
\label{t1t0}
\\
	\{ T^a[\a](s), T^b[\b](t) \} &=&
	\int du \dot\a^b(u) \d^3(\b(t),\a(u))
\nonumber
\\
	&& \quad
	(T^a[\b_t\!\circ_u\a] - T^a[\b_t\!\circ_{1-u}\a^{-1}])(\tilde u)
\nonumber
\\
	& - &
	\int du \dot\b^a(u) \d^3(\a(s),\b(u))
\nonumber
\\
	&& \quad
	(T^b[\a_s\!\circ_u\b] - T^b[\a_s\!\circ_{1-u}\b^{-1}])(\tilde u),
\eeqa
where $\tilde u$ is the parameter of the intersection of the combined loop.
Notice that the combination of loops occurs with a sign opposite to
(\ref{spinidsu2}) and cannot be further simplified. The variables $T^0$ and
$T^1$ are not expected to be canonical since they are of the type $x$ and
$xp$, respectively.

Initially, we define the space of states to be the space of functionals of
multiloops. The definition of the loop operators $\hat T^n$ can be
motivated via the Rovelli-Smolin transform (\ref{lt}) as explained in
section 3.2. We define
\beqa
	\hat T [\c] \psi[\eta] &=& \psi[\c\cup\eta],
\label{hatT0}
\\
	\hat T^a[\c](s) \psi[\eta] &=& - \hbar \int du \dot\eta^a(u)
	\d^3 (\c(s),\eta(u)) (\psi[\c_s\circ_u\eta]
	-\psi[\c_s\circ_{1-u}\eta^{-1}).
\label{hatT1}
\eeqa
The $\hat T[\c]$ operator acts by adding a loop into the multiloop argument
of the loop state. The $\hat T^a[\c](s)$ operator gives a non-zero result
only if the loop $\eta$ in the argument of the state intersects $\c$ at
$\c(s)$. In this case the result is a linear combination of the loop state
evaluated for the two different reroutings of the loop through the
intersection. In a graphical short hand,
\begin{center}
\mbox{\epsfig{figure=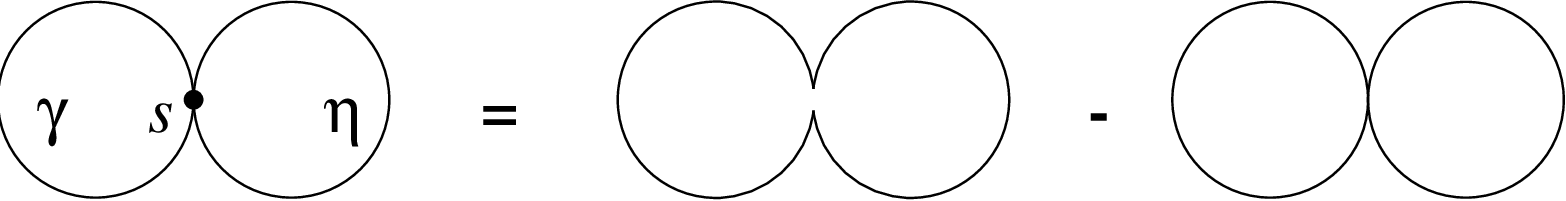,width=79mm}}
\end{center}
For $n\geq1$, $\hat T^{a\ldots b}[\c](s,\ldots,t) \psi[\eta]$
is non-zero only if the loops $\c$ and $\eta$ intersect in all the
distinguished points along $\c$, and the result is a linear combination of
all the possible reroutings of the two loops through the intersection.
For example, in the action of $\hat T^2$ there appear (with correct signs)
the four terms
\begin{center}
\mbox{\epsfig{figure=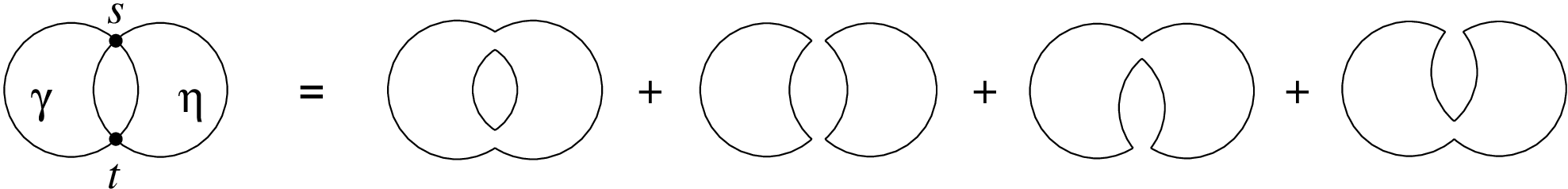,width=115mm}}
\end{center}

One can check that the relations $[\;,\;] = i\hbar \widehat{\{\;,\;\}}$,
(\ref{ccr}), are indeed satisfied for the operators $\hat T^0$ and $\hat
T^1$!  For the higher order loop operators one finds a deformation of the
type
\beq
	[\hat T^m, \hat T^n] \sim \hbar \hat T^{m+n-1} + \hbar^2 \hat
T^{m+n-2} + \ldots .
\eeq

Furthermore, we have to require that the action of the $\hat T^n$ is
consistent with identities for the $T^n$. This gives rise to several
properties of the loop states $\psi$. First of all, $\psi[\eta]$ has to be
invariant under reparametrizations of $\eta$. For any loops $\a$ and $\b$
and any path $\mu$, we require $\psi[\a\b]=\psi[\b\a]$ and
$\psi[\a\mu\mu^{-1}]=\psi[\a]$. In addition there are anti-commutation
relations (\ref{antic}) for the non-linear identity (\ref{spinidsu2})
where $\a$ and $\b$ intersect. We have for all $\eta$ that
\beqa
	(\hat T[\a] \hat T[\b] + \hat T[\b] \hat T[\a]) \psi[\eta]
&=&
	2 \psi[\a\cup\b\cup\eta],
\\
	2 \widehat{T[\a]T[\b]} \psi[\eta]
	= 2 (\hat T[\a\b] + \hat T[\a\b^{-1}]) \psi[\eta]
&=&
	2 (\psi[\a\b\cup\eta] + \psi[\a\b^{-1}\cup\eta]).
\eeqa
Hence we have to impose on the space of states the spinor identity
\beq
	\psi[\a\cup\b] =
	\psi[\a\b] + \psi[\a\b^{-1}].
\label{spinidpsi}
\eeq
As an immediate consequence we obtain that $\psi[\c_0\cup\eta] = 2
\psi[\eta]$ for $\c_0$ the zero loop,
and $\psi[\eta^{-1}]=\psi[\eta]$.

There also are identities among $T^0$ and $T^1$ variables (and similarly
for higher orders), but these
imply that the space of states contains only the zero state, $\psi\equiv0$
\cite{Br93a}. However, in a natural regularization of the $\hat
T^0$-$\hat T^1$ algebra (using strips, see below) such identities are
absent, and the spinor identity (\ref{spinidpsi}) is the only consequence
of the anticommutation relations.

\subsection{Constraints from Loop Variables}

How can we define the constraints, which are given to us in terms of the
Ashtekar variables, in terms of loop variables?  The basic observation is
that local variables in terms of $A$ and $E$ at a point $x$ can be obtained
from the non-local loop variables in the limit that a loop is shrunk to a
point. Considering that local and non-local field theories are
fundamentally different, the limit of shrinking loops is expected to be
non-trivial.

\begin{figure}
\par
\centerline{
\epsfig{figure=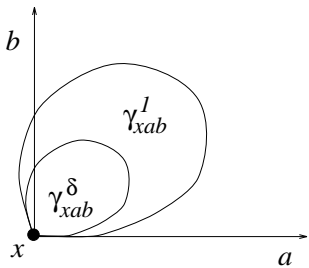,width=40mm}
}
\par
\caption{The family of small planar loops.}
\label{f6}
\end{figure}

We choose a coordinate system in a neighborhood of $x$, and we consider a
family of loops, $\c^\d_{xab}$, of size $\d$ at the point $x$ in the
$a$-$b$-coordinate plane (figure \ref{f6}) such that
\beqa
	\c^\d_{xab} (s) &=& x + \d \c_{ab}(s),
\\
	\c_{ab} (0) = \c_{ab} (1) &=& 0,
\\
	\sigma^{cd}(\c_{ab}) &=& \d^{[c}_a \d^{d]}_b,
\eeqa
where $\c_{ab}$ is a planar loop at the origin with unit area and area
element $\sigma^{cd}$, i.e. $\c^\d_{xab}$ has area $\d^2$. In the limit
$\d\rightarrow0$, $\c^\d_{xab}$ reduces to the zero loop, $\c^0_{xab}(s) =
x \:\forall s$.

The parallel transport around infinitesimal loops $\c^\d_{xab}$ gives rise
to the curvature $F$ of $A$ at $x$,
\beq
	U_{\c^\d_{xab}} = 1 + \d^2 F_{ab} (x) + O(\d^3),
\eeq
where $F^i_{ab} = \partial_aA_b^i - \partial_bA_a^i + \e^{ijk} A_a^j A_b^k$.
This is just the equation we need for the transition from
from the loop variables $T^1$ and $T^2$ to the constraints $D$ and $H$.

Indeed, in the Ashtekar variables the vector constraint $C$ and the
Hamiltonian constraint $H$ are (\cite{As91} or Giulini in this volume)
\beqa
	C_a(x) &=& \tr E^b F_{ab}(x),
\\
	H(x) &=& \tr E^a E^b F_{ab}(x).
\eeqa
The vector constraint generates diffeomorphisms up to gauge.
(As an aside, compare the canonical formulation in the metric variables,
e.g. Beig in this volume, where $H$ is more complicated, in particular due
to the presence of a potential term.)

In order to make the transition from a $T^a = \tr E^a U$ to $C_a = \tr E^b
F_{ab}$, we only have to shrink the loop. To be precise,
\beqa
	C_a(x) &=& \lim_{\d\rightarrow0} \frac{1}{\d^2}
	T^b[\c^\d_{xab}](0),
\\
	H(x) &=& \lim_{\d\rightarrow0} \frac{1}{\d^2}
	T^{ab}[\c^\d_{xab}](\d^2,1),
\eeqa
where the summation over indices is for $a\neq b$.

Quantization in the loop representation elevates the loop variables to the
loop operators, and since we know how to express the constraints in
the loop variables, we obtain immediately the constraints in the loop
representation via
\beqa
	C_a(T^1) &\rightarrow& \hat C_a = C_a(\hat T^1),
\\
	H(T^2) &\rightarrow& \hat H = H(\hat T^2).
\eeqa
{\em The question is whether the limit of shrinking loops can
actually be computed for the loop operators}, and with what result.

The natural differential operator in the context of shrinking loops turns
out to be the area derivative $\Delta_{ab}(s)$ of loop functionals. A loop
functional is called area differentiable if the following limit exists
independently of the choice of small loops $\c^\d_{xab}$ and transforms like
a two-form under coordinate transformations:
\beq
	\Delta_{ab}(s) \psi[\eta] = \lim_{\d\rightarrow0}
	\frac{\psi[\eta_s\!\circ\c^\d_{xab}] - \psi[\eta]}{\d^2},
\eeq
where the combination of the loops $\eta$ and $\c^\d_{xab}$ occurs at $x$.
Since we are considering general relativity, one has to show that a
background independent definition of the area derivative exists. This can
be shown for small parallelograms built from the integral curves of two
commuting vector fields \cite{BrPu93}. The beginning of a rigorous calculus
on the space of loop functionals can be found in the work of Tavares
\cite{Ta93}.

The area derivative satisfies
\beqa
	\Delta_{ab}(s) &=& \Delta_{[ab]} (s) ,
\label{ad1}
\\
	\frac{\d}{\d \eta^a(s)} &=& \dot\eta^b(s) \Delta_{ab}(s),
\label{ad2}
\\
	\Delta_{ab}(s) \tr U_\c &=&  \tr U_{\c_s^s} F_{ab}(\c(s)).
\label{ad3}
\eeqa
The area derivative is antisymmetric as expected from its definition in
terms of an area element (\ref{ad1}). The ordinary functional derivative
is a special, one-directional case of the area derivative (\ref{ad2}). And
the area derivative inserts $F$ into the Wilson loop as expected for the
limit of shrinking loops (\ref{ad3}).

\subsection{Constraint Operators}

In general terms, the area derivative is defined by a deformation of the
loop argument of a loop functional that consists of appending a small
loop. And the loop operators insert loops into the argument of loop
functionals in the specific way defined in (\ref{hatT0},\ref{hatT1}).
The perhaps surprising result is that the terms appearing in the
definition of the constraints can be combined into area
derivatives. In this calculation we have to make use of the spinor
identity for loop states, (\ref{spinidpsi}).

For the smeared diffeomorphism constraint $\hat D(v) = \int
d^3x v^a(x) C_a(x)$ we find that
\beqa
	\hat D (v) \psi[\eta] &\equiv&
	\lim_{\d\rightarrow0} \frac{1}{\d^2} 	\int d^3x v^a(x)
	T^b[\c^\d_{xab}](0)
\\
	&=&
	\int ds v^a(\eta(s)) \frac{\d}{\d\eta^a(s)} \psi[\eta],
\eeqa
which is the natural generator of diffeomorphisms along the vector field
$v^a$, and where we have used (\ref{ad2}).

%In the case of the Wheeler-DeWitt operator, one can separate the
%point-splitting procedure from the calculation that leads to the area
%derivative \ref{}.

\begin{figure}
\par
\centerline{
\epsfig{figure=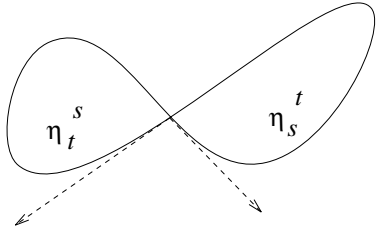,width=50mm}
}
\par
\caption{A loop with one self-intersection.}
\label{f7}
\end{figure}

The unregulated result for the smeared Hamiltonian
constraint $H(N) = \int d^3 N(x) H(x)$ is
\beq
	\hat H_{unreg} (N) \psi[\eta] =
	\int ds \int dt \d^3(\eta(s),\eta(t)) N(\eta(s)) \dot\eta^a(s)
	\dot\eta^b(t) \Delta_{ab}(s,\eta_t^s) \psi[\eta_t^s \cup \eta_s^t].
\label{hamilunreg}
\eeq
The second argument of the area derivative indicates on which of the
elements of the multiloop the area derivative acts. The result is non-zero
only if $\eta(s)=\eta(t)$. This occurs for $s=t$, but also for $s\neq t$ if
$\eta$ intersects itself. If $s=t$, then we expect a zero result since then
we have the antisymmetrized product of two identical tangent vectors
because the area derivative is antisymmetric in its indices, (\ref{ad1}).
If there is an intersection, then the action of $\hat H$ is to split the
loop $\eta$ at the intersection into a multiloop (figure \ref{f7}),
and to act with the area derivative on one of the component loops.

Notice that the Hamiltonian constraint operator has to be regulated. Since
the three-dimensional delta distribution $\d^3(\eta(s),\eta(t))$ depends on
only two parameters, there is a $\d^1(0)$ divergence. There are different
ways known to regulate the constraints, all of which share the same problem
that some sort of background dependence is introduced.

Let us comment on the point-splitting regularization most often used in
practice \cite{JaSm88,Br93c}. We pick a regulator $f_\e(x,y)$ such that
$\lim_{\e\rightarrow0} f_\e(x,y) = \d^3(x,y)$, and replace the delta
distribution by the regulator. The resulting operator, $\hat H_\e$, is
finite for $\e>0$. We perform all calculations for $\e>0$ to leading order
in $\e$, which is at order $1/\e$ and corresponds to $\d^1(0)$. We define
the regularized Hamiltonian constraint operator by a multiplicative
renormalization as $\hat H = \lim_{\e\rightarrow0} \e \hat H_\e$.

Background dependence enters since in $f_\e(x,y)$ we measure the
separation of $x$ and $y$ by a background metrice, e.g. a common choice
is $f_\e(x,y) = \frac{3}{4\pi\e^3} \Theta(\e - |x-y|)$. A background
dependence breaks diffeomorphism invariance and is therefore unacceptable
in quantum gravity. However, the leading order term in the Hamiltonian
constraint at intersections is background independent.
The $s=t$ terms vanish because of antisymmetry at order $1/\e^2$, but there
is a background dependent contribution at $1/\e$, which however vanishes
for diffeomorphism invariant states satisfying a certain regularity
conditions. Therefore, it is possible to discuss solutions to the
constraints in a background independent way, but there still remain
some problems regarding the regularization of the constraint algebra.

\subsection{Analytic Knot Invariants}

How can we find the simultaneous space of solutions to both the
diffeomorphism constraint and the Hamiltonian constraint of quantum
gravity? The problem is that very little is known about the Hamiltonian
constraint in the loop representation as a differential operator.  As a
first step we therefore consider a particular class of knot invariants and
examine the action of the Hamiltonian constraint on such knot invariants
\cite{BrGaPu92a}. At the time of writing, there exists only one class of
knot invariants for which we can compute the area derivative, namely the
analytic knot invariants that appear as coefficients of knot polynomials in
perturbative Chern-Simons theory.

The action of Chern-Simons theory is
\beq
	S_{CS}[A] = \int_\Sigma \!d^3x\: \e^{abc} \tr ( A_a\partial_bA_c +
	\frac{2}{3} A_aA_bA_c),
\eeq
where we choose for simplicity a $SU(2)$-valued one-form $A_a(x)$.  At this
point there is no relation between the Chern-Simons connection and the
Ashtekar connection, the Chern-Simons connection serves only as a
calculational tool. The vacuum expectation value of the Wilson loop in
Chern-Simons field theory is
\beq
	\langle h[\c,A] \rangle = \int\!DA\,h[\c,A] \exp
	\left(\frac{ik}{4\pi} S_{CS}[A]\right),
\eeq
where $k\in\integers$ is the coupling constant.

We have two ways to evaluate $\langle h[\c,A] \rangle$. Witten showed that
\beq
	\frac{\langle h[\c,A] \rangle} { \langle h[\c_0,A] \rangle }
	=
	c^{-w(\c)} J_q(\c)
\eeq
where $\c_0$ denotes the unknot, $q = \exp \frac{2\pi i}{k + 2}$, $c =
c(k)$, $w(\c)$ is the writhe of $\c$, and $J_q(\c)$ is the Jones
polynomial. The right hand side is also known as the Kauffman bracket
\cite{Ka91}.  The Jones polynomial is defined by the skein relations
(compare (\ref{skeinA})
\beq
	q J_q(c_+) - q^{-1} J_q(c_-) = (q^{\fot} - q^{-\fot}) J_q(c_=).
\eeq
The Jones polynomial is a knot invariant, or in knot diagram language, the
Jones polynomial is invariant under ambient isotopy.  The writhe $w(\c)$ is
defined as the sum over the crossings in a knot diagram counting $+1$ for
$c_+$ and $-1$ for $c_-$. The writhe is {\em not} a knot invariant, but
only a regular isotopy invariant (as is the Kauffman bracket).

\begin{figure}
\par
\centerline{
\epsfig{figure=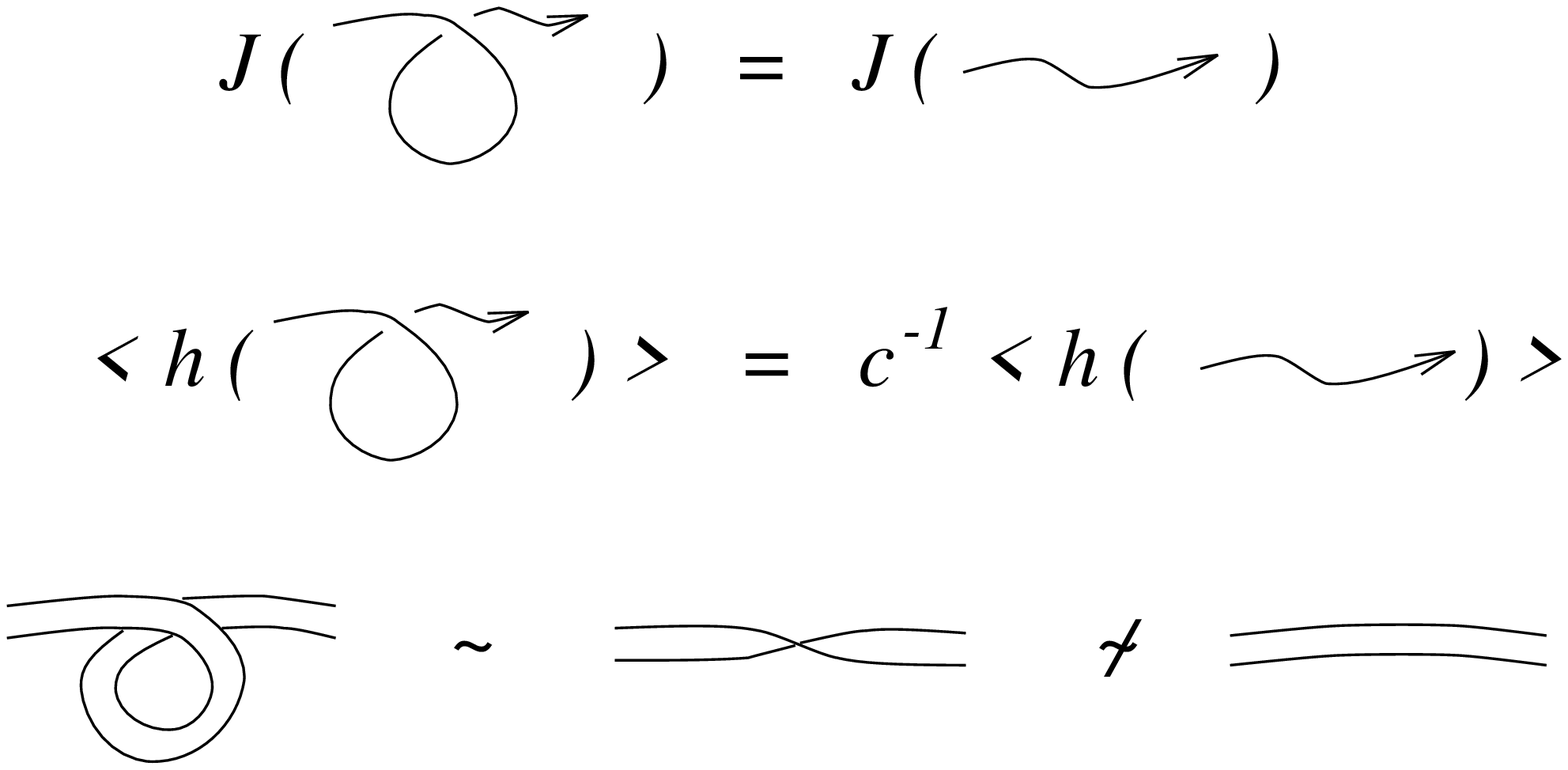,width=60mm}
}
\par
\caption{Framing dependence.}
\label{f8}
\end{figure}

The reason is that a projection of a knot from three into two dimensions may
introduce arbritrary numbers of crossings. For example, depending on the
projection we can obtain from an unknotted line segment a line without or
with twist, which are equivalent under ambient isotopy, but inequivalent
under regular isotopy. The corresponding crossing change formulas for
$J_q(\c)$ and $\langle h[\c,A] \rangle$ are shown in figure \ref{f8}. The
projection dependence of $ \langle h[\c,A] \rangle $ is known as framing
dependence. The framing of a loop can be defined in three dimensions by
replacing the loop by a strip, which changes under twists (figure \ref{f8}).

We can also obtain a perturbation expansion for $\langle h[\c,A] \rangle$
by inserting for $h[\c,A]$ the expansion (\ref{pexp}) \cite{GuMaMi90}. The
result is
\beq
	\frac{\langle h[\c,A] \rangle} { \langle h[\c_0,A] \rangle }
	=
	c_0(\c) + c_1(\c) \frac{1}{k} + c_2(\c) \frac{1}{k^2} + \ldots,
\eeq
where the coefficients $c_i(\c)$ are known to be regular isotopy invariants
related to the coefficients of the Jones polynomial by Witten's result. The
point is that the $c_i(\c)$ are expressed as multiple integrals along
the loop like the Gauss linking number $gl(\a,\b)$, (\ref{gl}). We have that
up to constant factors
\beqa
	c_0(\c) &=& 1,
\\
	c_1(\c) &=& gl(\c,\c),
\\
	c_2(\c) &=& (c_1(\c))^2 + \rho(\c).
\eeqa
$gl(\c,\c)$ is called the Gauss self linking number. Despite appearance,
$gl(\a,\b)$ in (\ref{gl}) is finite for $\a=\b$, but it depends on the
coordinates, which is another sign for framing dependence. If we assign to
$\c$ a framed loop $\c^f$, then we can define the framed self linking
number for the strip formed by $\c$ and $\c^f$ by $gsl^f(\c) = gl(\c,\c^f)$.
The second coefficient, however, contains a framing independent term
$\rho(\c)$, which is a true knot invariant related to the second
coefficient of the Alexander-Conway polynomial $a_2(\c)$, $\rho \sim a_2 +
\frac{1}{12} $. $\rho(\c)$ is the sum of a three-fold and a four-fold integral
along $\c$, but its precise form is not important here, since we are not
going to perform an explicit calculation with it. Each coefficient
$c_i(\c)$ can be shown to contain a framing independent piece, and these
constitute the class of knot invariants on which we want to study the
action of the Hamiltonian constraint.

\subsection{Knot Invariants as Solutions to the Constraints}

The history of solutions to the constraints in the loop representation
begins with the discovery in \cite{JaSm88} that in the connection
representation any $\psi[A] = h[\c,A]$ for some loop $\c$ without
intersections is a solution to the Wheeler-DeWitt equation. (Such states
are clearly not diffeomorphism invariant.) The simple reason is as in
(\ref{hamilunreg}) that two linearly independent tangent vectors are
required at one point for a non-zero result due to the presence of an
antisymmetrization.

\begin{figure}
\par
\centerline{
\epsfig{figure=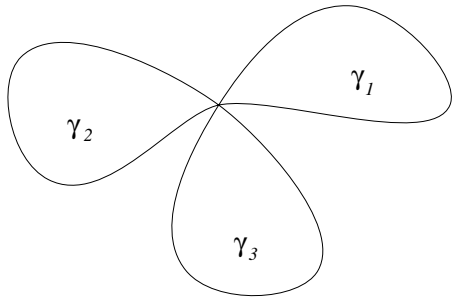,width=60mm}
}
\par
\caption{Generic loop with triple self-intersection and three linearly
independent tangents at the intersection (in three dimensions). }
\label{f9}
\end{figure}

In fact, the operator corresponding to the determinant of the three-metric,
$\widehat{\sqrt{- g}}$, gives zero if there are not three linearly
independent tangents at one point, which for smooth loops can only happen
at a triple intersection (figure \ref{f9}). Since the metric itself is not
an observable of the theory, one might think that one does not have to
worry about its degeneracy on the space of solutions. However, the
determinant of the metric typically appears in matter couplings, and it
should therefore not be zero. Also, the Hamiltonian for a non-zero
cosmological constant $\L$ is
\beq
	\hat H_\Lambda = \hat H + \Lambda \widehat{ \sqrt{- g}}.
\eeq
States without intersections are therefore solutions to the Wheeler-DeWitt
equation for arbitrary cosmological constant. We consider this as an
argument that the sector of states for non-intersecting loops is
degenerate.

Consequently, solutions to the Hamiltonian constraint for intersecting
loops were constructed in the connection representation
\cite{JaSm88,Hu89,BrPu91}. However, in \cite{BrPu91} it is shown that
all such solutions are necessarily annihilated by $\widehat{\sqrt{- g}}$.
Since in the connection representation we do not know a general strategy to
solve the diffeomorphism constraint anyway, it is natural to look for
non-degenrate solutions to the Hamiltonian constraint in the loop
representation. One reason why non-degenerate solutions can be found in the
loop representation is that the loop representation is based on the
opposite factor ordering of the $\hat A$ and $\hat E$ then the one in which
$h[\c,A]$ leads to solutions.

Let us summarize the situation in the loop representation for the
coefficients $c_i(\c)$ \cite{BrGaPu92a}.  $c_0$ is a non-degenerate
solution, but not an interesting one.  Let us consider loop states with
support on loops $\c = \c_1\c_2\c_3$, $\c_i$ without self-intersections
(figure \ref{f9}), which have one triple intersection with three linearly
independent tangent vectors (the generic case in three dimensions). We
find that
\begin{center}
\begin{tabular}{c|c|c}
	$\psi[\c]$ & $\hat H \psi[\c]$ & $\widehat{\sqrt{- g}} \psi[\c]$
\\ \hline
	$c_1(\c)$ & nonzero & nonzero
\\
	$c_1(\c_1\cup\c_2\cup\c_3)$ & 0 & 0
\\ \hline
	$\rho(\c)$  & 0 & nonzero!
\end{tabular}
\end{center}
$c_1$ (which is not a knot invariant), 	is non-degenerate but not
annihilated by $\hat H$. For the more symmetric loop as shown, $\c_1$ is
a degenerate solution. But a non-trivial and non-degenerate solution to
both constraints is the second coefficient of the Alexander-Conway
polynomial. In \cite{BaGaGrPu93} the same is shown for a part of $c_3$.

So, using a particular factor ordering and regularization, we can find
interesting solutions to both constraints among the coefficients $c_i(\c)$.
The mere fact that solutions can be found is a success of the loop
representation, since as already mentioned, in the traditional variables
not a single solution to the full Wheeler-DeWitt equation had been known.

However, it is still completely unclear what kind of structure the {\em
complete} space of solutions might have. The following argument provides at
least some insight into what this structure could be.  In the connection
representation there is exactly one solution known to all constraints
\cite{Ko90}, and that in the factor ordering that corresponds to the loop
representation and for non-vanishing cosmological constant:
\beq
	\psi_\Lambda[A] = \exp (-\frac{6}{\Lambda} S_{CS}[A]),
\eeq
where $A$ is now the Ashtekar connection. This state is gauge invariant for
appropriate values of $\Lambda$ and diffeomorphism invariant since
$S_{CS}[A]$ is 'topological'. Since
\beqa
	&& \hat H_\Lambda = \e^{ijk} \frac{\d}{\d A_a^i} \frac{\d}{\d A_b^j}
	F_{ab}^k + \frac{\Lambda}{6} \e^{ijk}\e_{abc}
	\frac{\d}{\d A_a^i} \frac{\d}{\d A_b^j} \frac{\d}{\d A_c^k} ,
\\
	&& \frac{\d}{\d A_c^k} \exp (-\frac{6}{\Lambda} S_{CS}[A]) =
	\frac{3}{\Lambda} \e^{cde} F_{de}^k \exp (-\frac{6}{\Lambda}
	S_{CS}[A]),
\label{dSCS}
\eeqa
we immediately have by differentiating only once that
\beq
	\hat H_\Lambda \psi_\Lambda[A] = 0.
\eeq
In other words, property (\ref{dSCS}) of the Chern-Simons action makes it
possible to choose the coefficient in the exponent of $\psi_\Lambda[A]$
such that the contributions to $\hat H_\Lambda$ from $\hat H$ and the
cosmological constant term cancel.

The idea in \cite{BrGaPu92b} is to consider the loop transform (\ref{lt}) of
$\psi_\Lambda[A]$,
\beq
	\psi_\Lambda[\c] = \int d\mu(A) h[\c,A] \psi_\Lambda[A].
\eeq
If the loop transform exists, then $\psi_\Lambda[\c]$ is a solution to all
the constraints in the loop representation by construction. In general, we
cannot compute the transform since we do not know the measure. Let us
assume for the moment, that for the transform of $\psi_\Lambda[A]$ we can
use the measure of Chern-Simons theory. Then
\beq
	\psi_\Lambda[\c] = \langle h[\c,A] \rangle_{CS} \sim c^{-w(\c)}
	J_q(\c),
\label{psil}
\eeq
for $k = -24/(\pi i\Lambda)$, is a solution to all constraints.

There are several obvious problems with this construction. In gravity, the
internal group is $SL(2,\complexes)$ and not $SU(2)$ (in terms of $A_a^i$,
in the former case $A_a^i$ is complex, in the latter it is
real). Furthermore, $\Lambda$ is complex (and takes discrete
values). Another problem is that we want to allow loops with intersections,
but we can in fact construct an extension of the Jones polynomial to
intersecting loops \cite{BrGaPu92b}.

While there is no proof that (\ref{psil}) does or does not make sense, it
hints at a very interesting relation between quantum gravity and
topological field theory. Furthermore, from the single solution
(\ref{psil}) for $\Lambda \neq 0$ we can derive a whole tower of solutions
for $\L = 0$ by the following simple argument \cite{BrGaPu93}. Consider
\beq
	\hat H_\L \psi_\L[\c] = (\hat H + \L \widehat{\sqrt{- g}}) (c_0 +
	c_1 \L + c_2 \L^2 + \ldots ) = 0,
\eeq
where we have absorbed the factor between $\L$ and $k$ in the $c_i$.
Since this equation holds for all $\L \neq 0$, we conclude order for order
that
\beqa
	\hat H c_0 &=& 0,
\\
	\hat H c_1 + \widehat{\sqrt{- g}} c_0 &=& 0,
\\
	\hat H c_2 + \widehat{\sqrt{- g}} c_1 &=& 0,
\\
	&\vdots& \nonumber
\eeqa
The action of $\widehat{\sqrt{- g}}$ is simpler to compute than that of
$\hat H$.  Since $c_2 = c_1^2 + a_2 + \frac{1}{12}$ and $\hat H c_1^2 +
\widehat{\sqrt{- g}} c_1 =0 $ we arrive at a simple proof of $\hat H
a_2 = 0$ for loops as above.

{}From the structure of the series, we can guess that at each order a part of
the coefficients $c_i$ has to be annihilated by $\hat H$.  Recently it has
been argued that indeed $\hat H c^{-w(\c)} = 0$ \cite{GaPu93}, and
therefore $\hat H J_{q(\L)} = 0$. This explains why each coefficient
contains a diffeomorphism invariant part that is a solution to the
Hamiltonian constraint (recall that $w(\c)$ is not diffeomorphism
invariant).

%%%%%%%%%%%%%%%%%%%%%%%%%%%%%%%%%%%%%%%%%%%%%%%%%%%%%%%%%%%%%%%%%%
\section{Discussion}

\subsection{Simple loop representations}

The main motivation for our discussion of the loop representation comes
from quantum gravity. Before we discuss the status of the loop
representation for quantum gravity in the next section, let us at least
briefly comment on the loop representation of simpler models.
\begin{enumerate}

\item Maxwell theory.\\
 As we have seen in section 3, the loop representation
is equivalent to the Fock representation and as complete. Here we can gain
some intuition about the physical meaning of loops.

\item Yang-Mills theory (see Loll in this volume).\\
As mentioned in section 2.3, loop variables play a natural role in
non-abelian gauge theories both on the lattice and in the continuum. The
mathematical problems that effectively stopped the continuum approach in
the beginning of the 1980's are now being adressed by results stimulated by
the loop representation of quantum gravity, see below. On the lattice, we
have a complete formulation, and it is well suited for numerical
Hamiltonian lattice gauge theory. For example, one can compute the glue
ball state and its mass in the high temperature regime of pure 2+1 $SU(2)$
lattice gauge theory, i.e. numerically solve the eigenvalue problem for the
Hamiltonian (here not a constraint), $\hat H \psi[\c] = E \psi[\c]$
\cite{Br91} (see also \cite{GaLeTr89}). We can introduce matter into the loop
representation by including paths that carry fermions at their ends. Again,
on the lattice we can perform numerical computation, see \cite{GaSe93} for
$SU(2)$ with fermions in 3+1 dimensions.

\item Linearized gravity. \\
Similar to the situation in Maxwell theory, the loop representation is able
to reproduce the Fock representation of gravitons
\cite{AsRoSm91}, and hence the elementary excitations of the gravitational
fields are based on loops. Furthermore, one can actually show that the loop
representation of quantum gravity, although incomplete, contains the
singular limit of gravitons \cite{IwRo93}.

\item 2+1 gravity. \\
The loop representation can be constructed explicitly \cite{AsHuRoSaSm89}.
Many of the problems regarding regularization are absent, notice for
example that in (\ref{hamilunreg}) there appears only $\d^2(\c(s),\c(t))$.
Together, the diffeomorphism and the Hamiltonian constraint generate
homotopy transformations (that can take intersections apart and transform
the unknot to the trefoil, figure \ref{f1}) \cite{BrVa93}. An important
point is that the loop transform in its naive form is degenerate
\cite{Ma93}, but there exists a non-degenerate generalization
\cite{AsLo93}.

\item Classical limit in the loop representation --- weaves. \\
On the kinematic level, i.e. ignoring the Hamiltonian constraint, so-called
weave states allow us to define a classical metric \cite{AsRoSm92}. A weave
is a multiloop obtained by sprinkling loops randomly into the
manifold. Under appropriate conditions, weave states are the eigenstates of
a smeared, diffeomorphism invariant metric operator, and the classical
metric arises as its eigenvalue on scales large in comparison to the
density of the sprinkling.  Remarkably, in this context we can {\em derive}
that physics becomes discrete at the Planck scale. Weaves can acommodate,
for example, the classical black hole solution \cite{Ze93}.

\end{enumerate}

\subsection{Status of the Loop Representations for Quantum Gravity}

We have seen in some detail how various steps of the program of algebraic
quantization of quantum gravity can be performed in the loop
representation. Let us collect the main negative and positive points about
this approach, the negative ones first:
\begin{description}

\item[--$\mbox{}^\infty$]
{\em The program is incomplete.}  As long as the program is incomplete, it
is totally unclear whether any parts of it will be part of the 'final'
theory of quantum gravity. It does not help that all other approaches to
quantum gravity are incomplete, too (see for example Isham in this
volume). Whenever the claim of progress in full quantum gravity is raised,
so far it is only valid with respect to a particular program.
More on the positive side, neither has it been shown that the loop
representation must necessarily fail.

\item[--]
The main reason why the loop representation is incomplete is that we do not
have an inner product, in particular we do not know how to obtain a
complete set of observables that could lead us to an inner product.
The absence of an inner product implies, for example, that all we have are
examples for $\psi[\c] \in V_{sol}$, not for $\psi[\c] \in {\cal H}_{phys}$.

\item[--]
The construction of the loop representation ignores the reality conditions.
The Ashtekar variables are complex, and certain reality conditions are
imposed to obtain real general relativity. We do not know how to impose the
reality conditions in the loop representation, although there may be an
analog to holomorphicity of $\psi[A]$ in the connection
representation. Also, the issue of reality becomes intertwined with that of
self-adjointness and the inner product. Therefore, at the level of our
discussion we deal with a quantum theory of complex general relativity.

\item[--]
Regularization is not complete. We know how to regularize the $\hat
T^0$-$\hat T^1$ algebra (using strips for $\hat T^1$, e.g. \cite{Ro91}),
and the constraints are reasonably well understood in terms of a
point-splitting regularization. What is missing is a regularization of the
full $\hat T^n$ algebra, and of the constraint algebra. Without progress on
regularization we cannot hope to decide, for example, whether there are
anomalies.

\item[--]
The loop representation has not lead to a breakthrough regarding the
interpretation of quantum gravity, say of observables or the issue
of time.

\end{description}
There are, however, also several positive aspects of the loop
representation approach:
\begin{description}

\item[+]
The loop representation is natural for the treatment of the constraints. As
argued in depth, the loop representation is well-adapted not only to gauge
and diffeomorphism invariance, but also the Wheeler-DeWitt equation seems
to become tractable in the loop representation.

\item[+]
There exist loop representations for many models that are simpler than full
quantum gravity in 3+1 dimensions, that can be called complete to a varying
degree (section 5.1). This fact gives us confidence that at least the general
idea behind the loop representation is sound. Of course, there are many
good ideas that cease to be valid in the case of quantum gravity.
Therefore the previous point is important.

\item[+]
There has been progress on several mathematical aspects related to loops:
\begin{enumerate}
\item
There are two, not in any obvious way related ways to extend the space of
loop states to distributions on the space of loops, \cite{AsIs92} and
\cite{BaGaGr93}. As usual in quantum field theory, we expect such states
rather than just functions on configuration space to be relevant.
\item
A differential calculus for extended loop variables has been developed
\cite{Ta93} that allows one to give rigorous meaning to heuristic
constructions of operators like the area derivative.
\item
A diffeomorphism invariant measure has been constructed on a completion of
the space of connections modulo gauge \cite{AsLe93}, see also \cite{Ba93}.
This measure may lead to a rigorous definition of the loop representation
via the transform.
\item
There are new knot invariants that are more powerful then the knot
polynomials, the Vassiliev invariants \cite{Va90}. These can be
characterized as the coefficients of knot polynomials (e.g. $a_2$
corresponds directly to a Vassiliev invariant), which they include as a
special case \cite{Bi93}. For the construction of the Vassiliev invariants,
intersecting loops are essential. Before the Vassiliev invariants became
known, knots with intersections were not studied in knot theory, but given
the importance of Vassiliev invariance in knot theory, the application of
loops with intersections in quantum gravity has gained additional
justification.
\end{enumerate}
\item[+]
There are new ideas about quantum gravity physics that have been
introduced, or concretized, by the loop representation. Although these
ideas refer mostly to limiting situations of full quantum gravity, this is
where our intuition originates, and if as often assumed new conceptual
ideas are needed for the quantization of gravity, it is good to know that
the loop representation produces such ideas. As mentioned in the context of
weaves \cite{AsRoSm92}, in the classical approximation to full quantum
gravity in the loop representation, the discrete structure of space time at
the Planck scale can be derived.  Furthermore, loops allow one to construct
diffeomorphism invariant observables that measure the area of a surface by
counting the number of intersections of a loop with this surface (which is
diffeomorphism invariant). A prediction of such a framework is that area is
quantized.
\end{description}

\subsection{Conclusion}

As it is often the case with a theory, it is a matter of taste and interests
whether one feels that the pro outweighs the contra --- especially with a
theory as remote from the 'real world' as quantum gravity. Let us draw our
conclusion.

We believe that the loop representation is an interesting proposal about
how to solve some of the long-standing problems of canonical quantum
gravity, in particular solving the constraint equations. While far from
complete, the loop representation offers promise for the future.

Here we have focused on three issues. For Maxwell theory we have argued
that in the loop representation we can replace the physical picture of
photons by that of elementary excitations based on loops, both being
equivalent. For full quantum gravity we have shown how to find states
that solve all the constraints in the loop representation,
something not possible in the traditional approach in terms of metric
variables.

The one aspect of the loop representation that arguably is the most important
one is the following. The loop representation is {\em not} a strange idea
unrelated to physics found in an appendix to an obscure theory called
canonical quantum gravity.  Rather the loop representation is a fine
example for the surprisingly fruitful interplay between three, initially
unrelated theories: knot theory, gauge theory, and quantum gravity.

\subsection*{Acknowledgements}
It is a pleasure to thank the organizers of the 117. Heraeus Seminar, and
especially Helmut Friedrich, for a well-organized and very interesting
meeting. This paper is based on the two lectures that the author presented
at the seminar. In my work on the loop representation I have greatly
benefitted from stimulating discussions with Abhay Ashtekar, Rodolfo
Gambini, Jorge Pullin, and Lee Smolin.

%%%%%%%%%%%%%%%%%%%%%%%%%%%%%%%%%%%%%%%%%%%%%%%%%%%%%%%%%%%%%%%%%%

\newcommand{\bib}[1]{\bibitem[#1]{#1}}
\newcommand{\cqg}[1]{{\em Class.\ Quan.\ Grav.\ }{\bf #1}}
\newcommand{\grg}[1]{{\em Gen.\ Rel.\ Grav.\ }{\bf #1}}
\newcommand{\np}[1]{{\em Nucl.\ Phys.\ }{\bf #1}}
\newcommand{\pr}[1]{{\em Phys.\ Rev.\ }{\bf #1}}
\newcommand{\prl}[1]{{\em Phys.\ Rev.\ Lett.\ }{\bf #1}}
\newcommand{\pl}[1]{{\em Phys.\ Lett.\ }{\bf #1}}
\newcommand{\jmp}[1]{{\em J. Math.\ Phys.\ }{\bf #1}}
\newcommand{\jgp}[1]{{\em J. Geom.\ Phys.\ }{\bf #1}}
\newcommand{\cmp}[1]{{\em Commun.\ Math.\ Phys.\ }{\bf #1}}
\newcommand{\mpl}[1]{{\em Mod.\ Phys.\ Lett.\ }{\bf #1}}
\newcommand{\ijmp}[1]{{\em Int.\ J. Mod.\ Phys.\ }{\bf #1}}
\newcommand{\apny}[1]{{\em Ann.\ Phys.\ (N.Y.) }{\bf #1}}


\begin{thebibliography}{BrGaPu92a}

\bib{}
No attempt was made to cover all the literature related to the loop
representation, but let us point out to the non-specialist some of the
references that can serve as entry points to the literature. For a guide to
the issues of quantum gravity in general see Isham in this volume.  For a
review of the loop representation in Yang-Mills theory see Loll in this
volume. For a list of references on canonical quantum gravity in the
Ashtekar variables in general and the loop representation in particular see
\cite{Br93b}. Taken together, the latter two sources give a rather complete
overview of all the work related to the loop representation.

For a complete and authorative review of quantum gravity in the Ashtekar
variables see the book by Ashtekar
\cite{As91}. There are also review papers by Rovelli \cite{Ro91} and Smolin
\cite{Sm93} that cover the loop representation. For a self-contained
introduction to the application of knot theory to quantum gravity along the
lines of section 4 see Pullin \cite{Pu93}. For a recent account of knot
theory in physics see Kauffman \cite{Ka91}.

\bib{Al28}
J.W.\ Alexander. Topological invariants of knots and links.
{\em Trans.\ Am.\ Math.\ Soc.\ }{\bf 30} (1928) 275--306

\bibitem[As86]{As86}
A. Ashtekar.
New variables for classical and quantum gravity.
\prl{57} (1986) 2244--7

\bibitem[As87]{As87}
A. Ashtekar.
New Hamiltonian formulation of general relativity.
\pr{D36} (1987) 1587--1602

\bibitem[As91]{As91}
A. Ashtekar. Lectures on non-perturbative canonical gravity.
(World Scientific, Singapore 1991)

\bib{AsHuRoSaSm89}
A. Ashtekar, V. Husain, C. Rovelli, J. Samuel, and L. Smolin.
$2+1$ quantum gravity as a toy model for the $3+1$ theory.
\cqg{6} (1989) L185--93

\bibitem[AsIs92]{AsIs92}
A. Ashtekar and C. Isham.
Representations of the holonomy algebras of gravity and non-abelian
gauge theories.
\cqg{9} (1992) 1433--85

\bib{AsLe93}
A. Ashtekar and J. Lewandowski. Representation theory of analytic holonomy
$C^*$ algebras.  In ``Knots and Quantum Gravity'', ed.\ J. Baez.
(Oxford U. Press, in press)

\bib{AsLo93}
A. Ashtekar and R. Loll. A new loop transform for 2+1 gravity. In
preparation.

\bibitem[AsRo92]{AsRo92}
A. Ashtekar and C. Rovelli.
A loop representation for the quantum Maxwell field.
\cqg{9} (1992) 1121--50

\bibitem[AsRoSm91]{AsRoSm91}
A. Ashtekar, C. Rovelli and L. Smolin.
Gravitons and loops.
\pr{D44} (1991) 1740--55

\bibitem[AsRoSm92]{AsRoSm92}
A. Ashtekar, C. Rovelli and L. Smolin.
Weaving a classical geometry with quantum threads.
\prl{69} (1992) 237--40

\bibitem[Ba93]{Ba93}
J.C.~Baez.
Diffeomorphism-invariant generalized measures on the space of
connections modulo gauge transformations.
To appear in the proceedings of the ``Conference on Quantum Topology'',
Manhattan, Kansas, March 1993

\bibitem[BaGaGr93]{BaGaGr93}
C. Di Bartolo, R. Gambini, and J. Griego.
The extended loop group: an infinite dimensional manifold associated
with the loop space. Montevideo preprint
IFFI/93.01

\bib{BaGaGrPu93}
C. Di Bartolo, R. Gambini, J. Griego, and J. Pullin.
In preparation.

\bib{Be74}
W. Berkson.
Fields of force: the development of a world view from Faraday to Einstein.
(Routledge \& Kegan Paul, London 1974)

\bib{Bi93}
J. Birman.
New points of view in knot theory.
{\em Bull.\ Am.\ Math.\ Soc.\ }{\bf 28} (1993) 253--87

\bibitem[Bl90]{Bl90}
M.~P. Blencowe.
 The Hamiltonian constraint in quantum gravity.
\np{B341} (1990) 213--51

\bib{Br91}
B. Br\"ugmann.
The method of loops applied to lattice gauge theory.
\pr{D43} (1991) 566--79

\bib{Br93a}
B. Br\"ugmann.
Ph.D.\ thesis. (Syracuse University, 1993)

\bib{Br93b}
Bibliography of publications related to classical and quantum gravity in
terms of the Ashtekar variables. MPI preprint MPI-Ph/93-68 (Sept.\ 1993)

\bib{Br93c}
B. Br\"ugmann.
The Wheeler-DeWitt operator in the loop representation. Unpublished.

\bib{BrGaPu92a}
B. Br\"{u}gmann, R. Gambini and J. Pullin.
Knot invariants as nondegenerate quantum geometries.
\prl{68} (1992) 431--4

\bibitem[BrGaPu92b]{BrGaPu92b}
B. Br\"{u}gmann, R. Gambini and J. Pullin.
Jones polynomials for intersecting knots as physical states of quantum
gravity.
\np{B385} (1992) 587--603

\bibitem[BrGaPu93]{BrGaPu93}
B. Br\"ugmann, R. Gambini and J. Pullin.
How the Jones polynomial gives rise to physical states of quantum
general relativity.
\grg{25} (1993) 1--6

\bibitem[BrPu91]{BrPu91}
B. Br\"ugmann and J. Pullin.
Intersecting N loop solutions of the Hamiltonian constraint
of quantum gravity.
\np{B363} (1991) 221--44

\bibitem[BrPu93]{BrPu93}
B. Br\"ugmann and J. Pullin.
On the constraints of quantum gravity in the loop representation.
\np{B390} (1993) 399--438

\bibitem[BrVa93]{BrVa93}
B. Br\"ugmann and M. Varadarajan.
Unpublished notes (1993)

\bib{Di65}
P. A. M. Dirac. Lectures on Quantum Mechanics. (Academic Press, NY 1965)

\bib{Ga1833}
C. F. Gauss.
Note of Jan.\ 22, 1833. In ``Werke'', vol.\ V (K\"onigliche Gesellschaft der
Wissenschaften, G\"ottingen, 1877) 605

\bibitem[Ga91]{Ga91}
R. Gambini.
Loop space representation of quantum general relativity and the group of loops.
\pl{B255} (1991) 180--8

\bib{GaLeTr89}
R. Gambini, L. Leal, and A. Trias.
Loop calculus for lattice gauge theories.
\pr{D39} (1989) 3127--35

\bib{GaPu93}
R. Gambini and J. Pullin.
The Gauss linking number in quantum gravity.
In ``Knots and Quantum Gravity'', ed.\ J. Baez.
(Oxford U. Press, in press)

\bib{GaSe93}
R. Gambini and L. Setaro. SU(2) QCD in the path representation. Montevideo
preprint (April 1993)

\bib{GaTr80}
R. Gambini and A. Trias.
Second quantization of the free electromagnetic field as quantum mechanics
in the loop space.
\pr{D22} (1980) 1380--4

\bib{GaTr83}
R. Gambini and A. Trias.
Chiral formulation of Yang-Mills equations: A geometric approach.
\pr{D27} (1983) 2935--39

\bib{GaTr86}
R. Gambini and A. Trias.
Gauge dynamics in the C-representation.
\np{B278} (1986) 436--48

\bib{Gi81}
R. Giles.
Reconstruction of gauge potentials from Wilson loops.
\pr{D24} (1981) 2160--8

\bibitem[GuMaMi90]{GuMaMi90}
E. Guadagnini, M. Martellini and M. Mintchev.
Wilson lines in Chern-Simons theory and link invariants.
\np{B330} (1990) 575--607

\bib{Ho51}
L. van Hove.
{\em Acad.\ Roy.\ Belg.\ Bull.\ Cl.\ Sci.\ }{\bf 37} (1951) 610

\bib{Hu89}
V. Husain.
Intersecting loop solutions of the Hamiltonian constraint of
quantum general relativity.
\np{B313} (1989) 711--24

\bib{IwRo93}
J. Iwasaki and C. Rovelli.
Gravitons from loops: non-perturbative loop-space quantum gravity contains
the graviton-physics approximation.
Pittsburgh preprint (April 1993)

\bibitem[JaSm88]{JaSm88}
T. Jacobson and L. Smolin.
Nonperturbative quantum geometries.
\np{B299} (1988) 295--345

\bib{Jo85}
V. Jones.
A polynomial invariant for links via von Neumann algebras.
{\em Bull.\ AMS} {\bf 12} (1985) 103--10

\bib{Ka91}
L. Kauffman. Knots and Physics. (World Scientific, 1991)

\bib{Ko90}
H. Kodama.
Holomorphic wavefunction of the universe.
\pr{D42} (1990) 2548--65

\bib{Lo91}
R. Loll.
A new quantum representation for canonical gravity and SU(2)
Yang-Mills theory.
\np{B350} (1991) 831--60

\bib{Ma62}
S. Mandelstam.
Quantum electrodynamics without potentials.
\apny{19} (1962) 1--24

\bibitem[Ma93]{Ma93}
D. Marolf.
Loop representations for 2+1 gravity on a torus.
Syracuse SU-GP-93/3-1;
An illustration of 2+1 gravity loop transform troubles.
gr-qc/9305015

\bib{Mi83}
A. Migdal.
Loop equations and $1/N$ expansion.
{Phys.\ Rep.\ }{\bf 102} (1983) 199--290

\bib{Po79}
A. M. Polyakov.
Gauge fields as rings of glue.
\np{B164} (1979) 171--88

\bib{Pu93}
J. Pullin.
Knot theory and quantum gravity: a primer.
In ``Proceedings of the Vth Mexican School of Particles and Fields'', ed.\
J. Lucio (World Scientific, 1993)

\bib{Re93}
A. Rendall.
Unique determination of an inner product by adjointness relations in
the algebra of quantum observables.
To appear in \cqg{}; erratum in preparation

\bibitem[Ro91]{Ro91}
C. Rovelli.
Ashtekar's formulation of general relativity and loop-space non-perturbative
quantum gravity: a report.
\cqg{8} (1991) 1613--75

\bibitem[RoSm88]{RoSm88}
C. Rovelli and L. Smolin.
Knot theory and quantum gravity.
\prl{61} (1988) 1155--8

\bibitem[RoSm90]{RoSm90}
C. Rovelli and L. Smolin.
Loop representation of quantum general relativity.
\np{B331} (1990) 80--152

\bib{RoTa89}
J. Romano and R. Tate.
Dirac versus reduced space quantisation of simple constraint systems.
\cqg{6} 1487--500

\bib{Sm89}
L. Smolin.
Invariants of links and critical points of the Chern-Simon path integral.
\mpl{A4} (1989) 1091--112

\bib{Sm93}
L. Smolin.
What can we learn from the study of non-perturbative quantum general
relativity?
Syracuse preprint (March 1993)

\bib{Ta1877}
P. G. Tait.
On knots I, II, III.
(Orig.\ publ.\ 1877, 1884, 1885, resp.)
In ``Scientific papers of P. G. Tait''
(Cambridge University Press, 1898) 273--347

\bibitem[Ta93]{Ta93}
J. Tavares.
Chen integrals, generalized loops and loop calculus.
Preprint, U. Porto (April 1993)

\bib{Th1869}
W. H. Thomson.
On vortex motion.
{\em Trans.\ R. Soc.\ Edin.\ }{\bf 25} (1869) 217--60

\bibitem[Va90]{Va90}
V.A.\ Vassiliev.
Cohomology of knot spaces. In ``Theory of singularities and
its applications'' (V. Arnold, ed.) {\em Advances in Soviet Math.} {\bf 1}
(1990) 23--69

\bib{Wi74}
K. Wilson.
Confinement of quarks.
\pr{D10} (1974) 2445--59

\bib{Wi89}
E. Witten.
Quantum field theory and the Jones polynomial.
\cmp{121} (1989) 351--99

\bib{Ze93}
J. Zegwaard.
The weaving of curved geometries.
\pl{B300} (1993) 217--22


\end{thebibliography}
\end{document}